\documentclass[11pt,letterpaper]{article}
\usepackage{fancybox}
\usepackage[utf8]{inputenc}
\usepackage{epsfig,graphicx}
\usepackage{multicol,pst-plot}
\usepackage{pstricks}
\usepackage{amsmath}
\usepackage{amsfonts}
\usepackage{soul}
\usepackage[toc,page]{appendix}
\usepackage{authblk}
\usepackage[left=2.54cm,right=2.54cm,top=2.54cm,bottom=2.54cm]{geometry}
\usepackage{bm}
\usepackage[english]{babel}
\usepackage[colorlinks]{hyperref}
\usepackage{cancel}
\usepackage{caption}
\usepackage{float}
\usepackage{upgreek}
\usepackage{subfigure}
\usepackage{siunitx}
\usepackage{color}
\usepackage{tikz}
\usepackage{listings}
\usepackage{minted}
\usepackage{mdframed}
\usepackage{multicol}
\renewcommand{\red}[1]{#1}

\definecolor{codegreen}{rgb}{0,0.6,0}
\definecolor{codegray}{rgb}{0.5,0.5,0.5}
\definecolor{backcolour}{rgb}{0.95,0.95,0.95}
\hypersetup{colorlinks=true,linkcolor=codegreen,citecolor=blue,filecolor=blue,urlcolor=magenta,}

\title{Activation Functions, Statistics and Learning of Higher-Order Interactions in Restricted Boltzmann Machines}
\author[1,*]{G. di Sarra}
\author[2]{Y. Roudi}
\affil[1]{Kavli Institute for Systems Neuroscience, Norwegian University of Science and Technology, 7491 Trondheim, Norway}
\affil[2]{Department of Mathematics, King's College London, London, Strand, WC2R 2LS, UK}

\begin{document}

\maketitle

\begin{abstract}
    The great success of neural networks primarily arises from the presence of the large number of weight parameters combined with nonlinearities in the input-output relationship of single neurons. In this work, we study the relationship between the statistical properties of the weights and the nonlinearity of the hidden unit in Restricted Boltzmann Machines (RBMs) on the one side, and the distribution induced on binary visible units. We do this for four commonly used activation functions: Linear, Step, ReLU, and Exponential, and make qualitative predictions about the ability of these models to learn distributions with strong higher order interactions over the visible nodes. We show that in general, in an ensemble of RBMs with Gaussian weights, these distributions are rare and hard to learn, except when the hidden unit activation function is an Exponential.
\end{abstract}

\section{Introduction}
Neural networks are comprised of groups of units, each performing a non-linear transformation on the input that they receive from other units. The strength of the input from one unit to another is determined by both the output of the former and the weights connecting the two. A major part of neural networks research then focuses on developing algorithms for adjusting these weights - the learning algorithm- so that the network performs a given task, e.g. generate outputs with a given statistics, or categorize a set of inputs \cite{allen2019learning, arora2019fine, allen2019convergence, chatterjee2022generalization, oymak2020toward,li2018learning}. When it comes to the form of non-linearity of the units,
theoretical work on attractor neural networks has shown that the spin-glass phase, an obstacle to successful retrieval, is significantly suppressed in networks of continuous valued activation function compared to its binary counterpart \cite{PhysRevA.42.7459, PhysRevA.43.2084,PhysRevA.42.2418,PhysRevE.73.061904,Treves_1990}
and that they can can operate closer to the Gardner storage capacity \cite{PhysRevLett.126.018301}. The role of activation functions in learning come from shallow supervised student-teacher settings \cite{oostwal2021hidden,citton2025phase,nishiyama2025solution,MANZAN2025}. It is well known that the use of ReLU units improves the convergence of learning compared to sigmoidal units, and that ReLU activation function of hidden units has been shown to enhance training and performance in supervised and unsupervised deep rectified networks \cite{pmlr-v15-glorot11a}. The Parametric ReLU (PReLU), a variant of ReLU, enabled models to surpass human performance in ImageNet classification \cite{7410480}. While all these studies emphasize the role of activation function non-linearity in the performance of various neural network architecture, they mostly focus on cases of supervised learning or biological models. Few studies focus on how activation functions can influence the properties of generative artificial neural networks such as the RBMs. In this paper, our aim is to address this issue by focusing on the activation function of the hidden layer of RBMs with binary visible units.\\
RBMs are two-layer bipartite neural networks capable of learning from raw, unlabeled data using efficient algorithms \cite{10.1007/978-3-642-33275-3_2,ACKLEY1985147}. State-of-the-art deep neural network architectures are much more complex and significantly more superior in performance compared to RBMs. However, RBMs remain of great value due to their theoretical interpretability, still maintaining a reasonable degree of performance in many tasks \cite{diSarra_2025}. Furthermore, the tractability of RBMs combined with high representational capacity -- their binary version is a universal approximator \cite{le2008representational}-- makes them ideal for improving our current understanding of the effects of single unit input-output functions.\\
The effects of changing hidden units' activation functions have been empirically studied in RBMs. For example, RBMs with ReLU activation functions showed improved training performance compared to RBMs with binary units \cite{10.5555/3104322.3104425}. On the theoretical side, statistical physics approaches to RBMs have provided insight into training modalities and phase diagrams of different RBM architectures \cite{decelle2021restricted, e23010034, Bonnaire2025}. Learning in linear RBMs has been linked to memory retrieval in pairwise Hopfield networks \cite{Barra:2012aa,fachechi2025fundamental}. Introducing nonlinear activation functions enables higher-order interactions between visible units and these interaction terms can be explicitly computed \cite{10.1162/neco_a_01420, decelle2024inferring}. Interpolating between the linear and sigmoidal cases, this linear-nonlinear transition has been used to map the RBM phase diagram \cite{PhysRevE.96.042156, PhysRevE.97.022310}. RBMs with ReLU hidden unit activation have been further explored, identifying distinct operational phases, including a \textit{compositional phase} where RBMs achieve optimal performance \cite{PhysRevLett.118.138301}. In this phase, visible patterns arise from the combination of a large but finite set of features encoded by strongly activated hidden units.\\
In the case of RBMs with binary visible units, the marginal distribution over visible nodes can be written as a sum of terms involving $I^{(s)}_{i_1,\dots i_s} v_{i_1}\dots v_{i_s}$  ({\it vide infra}). The interactions $I^{(s)}_{i_1,\dots i_k}$ can be analytically expressed in terms of the nonlinearity of the hidden layer and the weights connecting hidden and visible units \cite{10.1162/neco_a_01420}. Similar expression are also  known to exist for the Potts-binary RBM \cite{decelle2024inferring}. Using these analytical results, we investigate how different choices of nonlinear activation functions in the hidden layer of an RBM influence its ability to represent statistical regularities in the data. Starting from the interaction expressions \cite{10.1162/neco_a_01420}, we analytically compute the moments of the distribution of the resulting interactions of order $s$, $I^{(s)}_{i_1,\dots i_s}$, when the weights are drawn from a Gaussian distribution. We find that, overall, the resulting RBMs are low-order interaction models: stronger lower-order interactions relative to higher-order ones can be favored. We do, however, also find exceptions. Specifically, in an ensemble of random Gaussian RBMs, an Exponential activation function is much more likely to produce stronger higher-order interactions compared to ReLU or sigmoid transfer functions. This suggests that RBMs trained on data with high-order interaction are more likely to lead to RBMs with high-order interactions if the activation function is Exponential rather than e.g. ReLU or Sigmoid.

This paper is organized as follows: In Section 2 below, we go through the preliminary aspects of this study, including the definition of a RBM, what we mean by the hidden layer activation function, and how the model is mapped to an interaction model \cite{10.1162/neco_a_01420}. In Section 3, we compute the moments of the interaction distributions for every order across four activation functions: Linear, Step, ReLU, and Exponential. Furthermore, we show the consequences on the interactions landscape. In Section 4 we define \textit{decaying} and \textit{non-decaying} interaction models and we show numerical results by training RBMs on lattice gas models and random ground truth RBMs, and compare the resulting models with the analytical expressions. Finally, in the last Section, we discuss the implications and potential future developments of this study.

\section{Restricted Boltzmann Machines}
\begin{figure}[htbp]
\centering
\includegraphics[width=9cm]{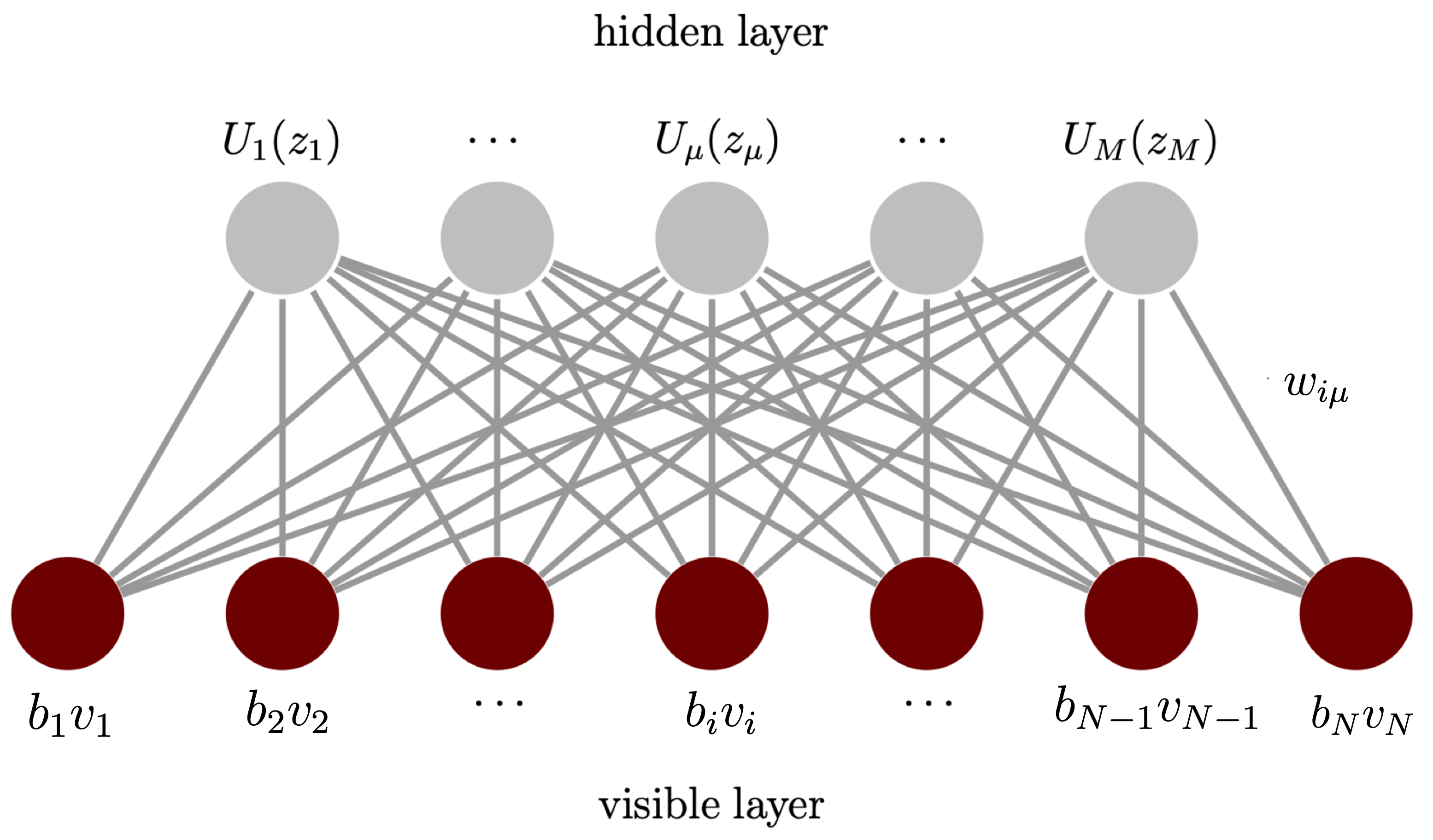}
\caption{Bipartite structure of a Restricted Boltzmann Machine.}
\label{fig:1}
\end{figure}

The Restricted Boltzmann Machine (RBM) is a two-layer stochastic neural network with 
$N$ visible and $M$ hidden units that we denote by $\bm{v} \equiv \{v_i\}, i = 1,\dots N$, and $\bm{z}\equiv \{z_{\mu}\}, \mu= 1\dots M$. The units are organized on a bipartite graph as shown in Figure \ref{fig:1}.

The joint distribution of the hidden and visible units is 
\begin{eqnarray}
    p(\boldsymbol{v}, \boldsymbol{z}) &=&\frac{1}{Z}\exp\biggl[-E(\boldsymbol{v}, \boldsymbol{z})\biggr],\\
    E(\boldsymbol{v},\boldsymbol{z}) &=& -\sum_{i}^{N}b_{i}v_{i} - \sum_{i,\mu}^{N,M}v_{i}w_{i\mu}z_{\mu} +\sum_{\mu}^{M}U_{\mu}(z_{\mu}),
    \label{eqref:p}
\end{eqnarray}
where $Z$ is the normalization constant, or partition function, $w_{i\mu}$ is the entry of the $N \times M$ weight matrix connecting visible unit $i$ and hidden unit $\mu$ and $b_{i}$ are the visible biases. As shown e.g. in \cite{10.1162/neco_a_01420,diSarra_2025} and explained in more details below, $U_{\mu}(z_{\mu})$ is a potential that defines the hidden layer activation function. In what follows, we use roman subscripts (e.g., $v_{i}$, $b_{i}$) to refer to variables associated to the visible units (e.g. activity, bias), and Greek subscripts (e.g., $z_{\mu}$, $c_{\mu}$) for the hidden ones.

In the classical formulation of the RBM the visible single units are $\{0,1\}$ binary variables. By choosing a linear potential $U_{\mu}(z_{\mu})= c_{\mu} z_{\mu}$ for $z_{\mu}=0,1$ and $U_{\mu}(z_{\mu})= +\infty$ otherwise, one can show that $p(z_{\mu}|h_{\mu})\propto \exp\left(z_{\mu}h_{\mu}\right)$ for $z_{\mu}=0,1$ and $p(z_{\mu}|h_{\mu})=0$ otherwise, where $h_{\mu}=c_{\mu}+\sum_{\mu}w_{i\mu}v_i$ is the input to hidden node $\mu$. The mode of $p(z_{\mu}|h_{\mu})$ is then $\hat{z}_{\mu}=\Theta(h_{\mu})$ where $\Theta(\cdot)$ is the Heaviside function and the mean has a sigmoidal relationship to $h_{\mu}$. The functional form that relates this mode or mean to $h_{\mu}$ is denoted as the activation function of the hidden units, in this case a Step or a sigmoid function. Table \ref{table1} shows the relationships between the hidden activation function thus defined, hidden potential and the mode or mean of $p(z_{\mu}|\boldsymbol{v})$ for four choices of the potential $U_{\mu}$.
\begin{table}[htbp]
\centering
\begin{tabular}{|c|l|c|c|}
\hline
\multicolumn{1}{|c|}{\textbf{Activation function}} & \multicolumn{1}{|c|}{\textbf{Hidden potential}} & \multicolumn{1}{|c|}{ $\bm{\hat{z}_{\mu}}$ } & \multicolumn{1}{|c|}{ $\bm{\tilde{z}_{\mu}}$} \\ \hline
Linear &
$ U_{\mu}= z_{\mu}^2/2+ c_{\mu}z_{\mu}$  &  $h_{\mu}$ & $h_{\mu}$ \\ \hline
Exponential & 
$ U_{\mu}=
\begin{cases}
  c_{\mu}z_{\mu} +\log z_{\mu}! & \text{if } z_{\mu} \in \mathbb{N} \\
  +\infty & \text{otherwise }
\end{cases} 
$   & $\text{floor}(\exp h_{\mu})$ & $\exp h_{\mu}$\\ \hline
Step &
$ U_{\mu}=
\begin{cases}
  c_{\mu} z_{\mu} & \text{if } \; z_{\mu}=0,1 \\
  +\infty  & \text{otherwise}
\end{cases} 
$  & $\Theta(h_{\mu})$ & $\frac{\exp h_{\mu}}{1+ \exp{h_{\mu}}}$ \\ \hline
ReLU & 
$ U_{\mu}=
\begin{cases}
  z_{\mu}^2/2+ c_{\mu}z_{\mu} & \text{if }\; z_{\mu} \geq 0 \\
  +\infty  & \text{if } z_{\mu} < 0
\end{cases} 
$  & $\text{max}(0,h_{\mu})$ & $h_{\mu}+\sqrt{\frac{2}{\pi}}\frac{\exp\left( \frac{-h_{\mu}^2}{2}\right)}{1+\text{erf}\left(\frac{h_{\mu}}{\sqrt{2}}\right)}$ \\ \hline
\end{tabular}
\caption{\textbf{Hidden unit activation functions.} For each activation function, the conditional probability $p(z_{\mu}|\boldsymbol{v})$ can be computed from Eq.~\eqref{eqref:p} with the hidden potential $U_{\mu}$ in the second column. The mode $\hat{z}_{\mu}$ and the mean $\tilde{z}_{\mu}$ of $p(z_{\mu}|\boldsymbol{v})$ are reported in the third and fourth column, where the input to hidden unit $\mu$ is $h_{\mu}=c_{\mu}+\sum_{\mu}w_{i\mu}v_i$. The $\text{floor}(\cdot)$ function computes the smaller closest integer of its argument and $\Theta(\cdot)$ is the Heaviside function.}
\label{table1}
\end{table}
Given these definitions of activation functions, in this work we will denote each potential and the corresponding activation function interchangeably: Gaussian and Linear, Step and Sigmoidal, Poisson and Exponential.

\subsection{RBMs as models of interacting variables}
\begin{figure}[htbp]
\centering
\includegraphics[width=\textwidth]{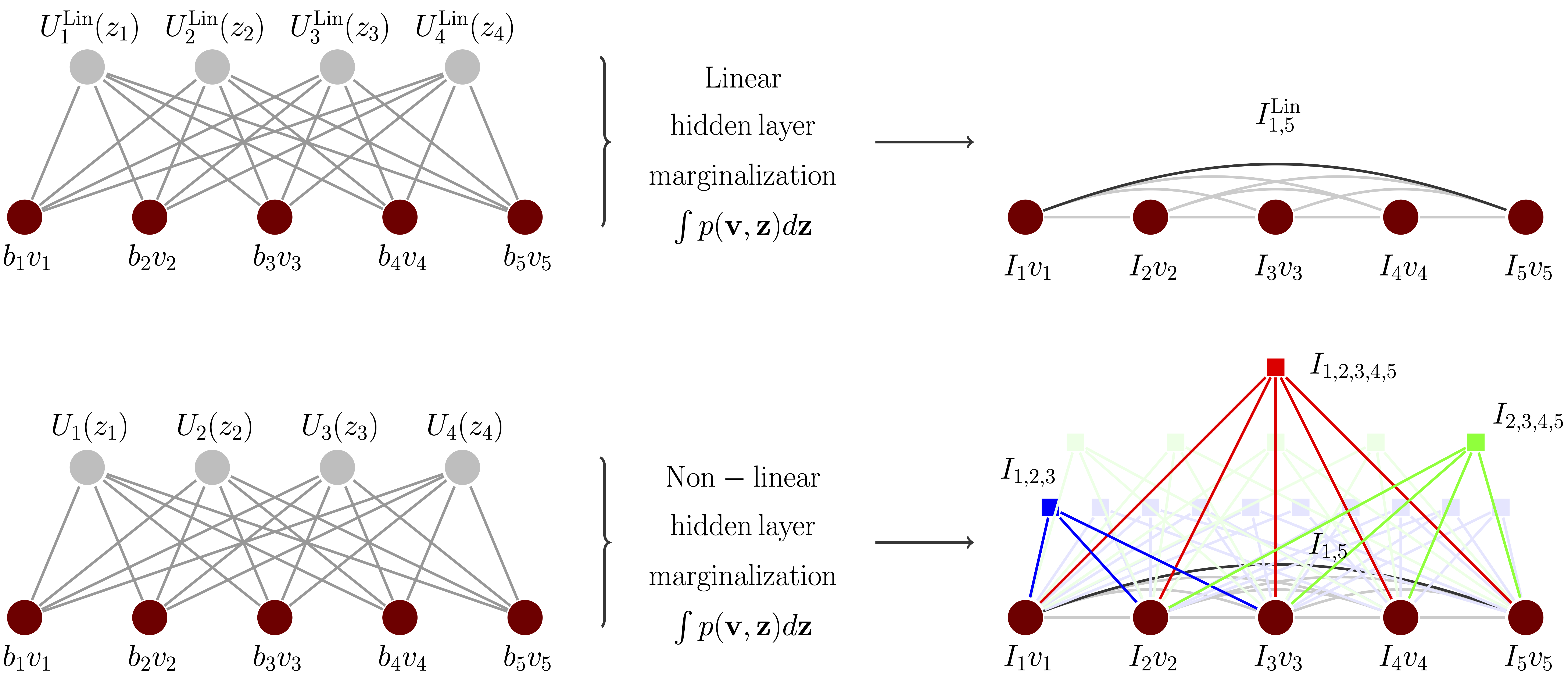}
\caption{Hidden layer marginalization. The joint distribution of an RBM with $N=5$ is marginalized with respect to the hidden layer to generate a fully-visible network with arbitrary orders of interaction between nodes. In the Linear RBM case, Eq. \eqref{marginal} corresponds to a Hopfield-like pairwise model. In the non linear cases, Eq. \eqref{marginal} also includes every higher-order interaction term up to $s=N$. The three-body interactions are represented in blue (highlighted $I_{1,2,3}$), the 4-body interactions in green (highlighted $I_{2,3,4,5}$), and the 5-body interaction $I_{1,2,3,4,5}$ in red.}
\label{fig:rbm_marg}
\end{figure}

For the model defined through Eq.~\eqref{eqref:p}, the marginal distribution for the visible units can be expressed as
\begin{equation*}
p(\boldsymbol{v}) =\frac{1}{Z} \exp \biggl[\sum_i b_i v_i + \sum_\mu K_{\mu}\biggl( \sum_i w_{i\mu}v_i\biggr)\biggr]
\end{equation*}
where $K_{\mu}(q_{\mu})= \log \mathbb{E}\left[ \exp \left(z_{\mu} q_{\mu}\right) \right]_{\rho(z_{\mu})}$ is the cumulant generating function of $\rho(z_{\mu})\propto \exp\left(-U_{\mu}(z_{\mu})\right)$.
This marginal can then be written as a model of interacting variables \cite{10.1162/neco_a_01420}, as
\begin{equation}
p(\boldsymbol{v}) =\frac{1}{Z'} \exp \left[ \sum_{k_1}I_{k_1}v_{k_1} + \sum_{k_1<k_2} I_{k_1,k_2}v_{k_1}v_{k_2}+...+I_{1,2,...,N}\prod_{k=1}^N v_k\right]
\label{marginal}
\end{equation}
A conceptual schematic describing the mapping of an RBM to the corresponding interacting variables model desribed above is depicted in Figure~\ref{fig:rbm_marg}.

Each of the terms in the exponent of Eq.~\eqref{marginal} takes the form $I_{i_1,\dots i_s} v_{i_1}\dots v_{i_s}$. When the nodes, $i_1 \dots i_s$, are all active, this term adds a value equal to $I_{i_1,\dots i_s}$ to the log of the marginal distribution. We refer to $I_{i_1,\dots i_s}$ as the “interaction term”, and as shown in \cite{10.1162/neco_a_01420}, they can be expressed as
\begin{equation}
I_{k_{1},...,k_{s}}=\sum_{\mu}^{M}\sum_{p=0}^{s-1}(-1)^{p}\sum_{j_{1}<j_{2}<...<j_{s-p}=1}^{s}K_{\mu}\left(\sum_{l=1}^{s-p}w_{k_{j_{l}}\mu}\right) + b_{k_1} \delta_{1,s}.
\label{exact}
\end{equation}
where $\delta_{s,s'}$ is the Kronecker function, while we refer to $w_{i\mu}, b_i$ and $c_{\mu}$ as the parameters of the RBM.

In particular, when $s = 2$, Eq.~\eqref{exact} represents the common case of pairwise interactions. If the hidden unit activation function is Linear (that is when $U_{\mu}$ is quadratic; see Table \ref{table1}), Eq.~\eqref{exact} takes a form akin to the Hebbian learning rule of the Hopfield model \cite{Barra:2012aa, 10.1162/neco_a_01420}.
The general form in Eq.~\eqref{exact} extends previous work on binary-linear RBMs \cite{Barra:2012aa} to binary-nonlinear RBMs with different activation functions for the hidden units, introducing interactions beyond pairwise. For specifications on the functional form of $K_{\mu}$ for different activation functions we refer to \cite{10.1162/neco_a_01420}.

\section{Statistics of induced interactions}
\label{sec3}
Our goal in this section is to compute the expectations for interactions of different orders and their correlations when the parameters are drawn from a given distribution. We define the expected value of a generic quantity $O$ over an ensemble of RBMs as
\begin{equation*}
    \left\langle O \right\rangle \equiv \int \prod_{k,\mu} d w_{k\mu} \int \prod_{\nu} d c_{\nu} P(\{w_{k\mu}\},\{c_{\nu}\})\ O(\{w_{k\mu}\},\{c_{\nu}\}),
\end{equation*}
where $P(\{w_{k\mu}\},\{c_{\nu}\} )$ is the distribution of the parameters. In what follows, unless otherwise stated, we assume that the parameters of the RBMs are independently and identically distributed according to $p(w_{k\mu})$ and $p(c_{\mu})$, that is
\begin{equation*}
P(\{w_{k\mu}\},\{c_{\nu}\})= \prod_{k,\mu} p(w_{k\mu}) \prod_{\nu} p(c_{\nu}),
\end{equation*}
and that $p(c_{\nu})= \delta(c_{\nu}-c_0)$, where $\delta(\cdot)$ is the Dirac $\delta$ distribution. 

\subsection{Exact interaction moments}
\label{3.1}
In the Linear and the Exponential cases, the relevant expectations  can be computed exactly. In the case of the Linear activation function, we recall that the cumulant generating function writes as  
\begin{equation*}
K^{\rm Lin}_{\mu}(q_{\mu}) = \frac{q^2_{\mu}}{2}-q_{\mu} c_{\mu},
\end{equation*}
and that induced interactions are non-zero only up to order two \cite{10.1162/neco_a_01420}:
\begin{equation}
\begin{aligned}
I^{(1)}_{k_i} = b_{k_i} +\sum_{\mu} \frac{{w_{k_i\mu}}^2}{2}-c_{\mu} w_{k_i\mu}, \qquad \qquad 
I^{(2)}_{k_i,k_j} =\sum_{\mu} w_{k_i\mu}w_{k_j\mu}\, .
\end{aligned}
\end{equation} 
The expected values of the induced interactions in this case are then 
\begin{equation}
\begin{aligned}
I^{\rm Lin}_1 &\equiv \langle I_{k_i} \rangle 
= b_{k_i} + \frac{M}{2}(w_0^2 + \sigma^2) - M w_0 c_0,\\
I^{\rm Lin}_2 &\equiv \left\langle I^{\rm Lin}_{k_1,k_2} \right\rangle 
= M w_0^2 \, .
\end{aligned}
\label{eq:lin_exact}
\end{equation}
where $w_0\equiv\langle w_{i\mu}\rangle$ , $\sigma^2 \equiv \langle w^2_{j\mu}\rangle - w^2_0 $ and M is the number of hidden nodes (see also Appendix \ref{appA}). Similarly, the second moments of the second order interaction are
\begin{align*}
&\left \langle {I^{\rm Lin}_{k_1,k_2}}^2 \right \rangle - \left \langle I^{\rm Lin}_{k_1,k_2} \right \rangle^2  = 2M \sigma^2(w_0^2 + \sigma^2/2),\\
&\left \langle I^{\rm Lin}_{k_1,k_2} I^{\rm Lin}_{k_1,k_3} \right \rangle - \left \langle I^{\rm Lin}_{k_1,k_2}\rangle \langle I^{\rm Lin}_{k_1,k_3} \right \rangle = Mw_0^2 \sigma^2, k_1\neq k_3.
\end{align*}
For all other cases, the correlations trivially factorize.

We can define the interaction terms dispersion as as the fluctuations in pairwise interactions relative to their mean, that is
\begin{equation*}
\Delta^{\rm Lin}_{k_1,k_2} \equiv \frac{\left \langle {I^{\rm Lin}_{k_1,k_2}}^2 \right \rangle - \left \langle I^{\rm Lin}_{k_1,k_2} \right \rangle^2}{\left \langle I^{\rm Lin}_{k_1,k_2} \right \rangle^2} = \frac{\sigma^2}{I^{\rm Lin}_{2}}\left [ 2+\frac{\sigma^2}{I^{\rm Lin}_{2}}\right] = \frac{u_{\rm Lin}}{M}\left(2+u_{\rm Lin}\right) 
\end{equation*}
where $u_{\rm Lin}\equiv\sigma^2/w_0^2$. Although very simple to derive, let us take a moment to note some properties of these equations. Firstly, as $M$ increases, for fixed $\sigma^2$ and $w_0 \neq 0$, the pairwise interactions and mean inputs increase linearly with $M$, but the dispersion $\Delta^{\rm Lin}_{k_1,k_2}$ decreases with $M$. When $w_0\to 0$, $I^{\rm Lin}_2$ goes to zero as $w_0$ and $\Delta^{\rm Lin}_{k_1,k_2}$ diverges as $w^{-4}_0$. Finally, while $I^{\rm Lin}_1$ depends on the second moment of the distribution of the weights and on the expected value of $c_{\mu}$, $I^{\rm Lin}_{2}$ depends only on the mean value of the weight distribution. All these properties are direct consequences of the simple form of the pairwise interactions for the Linear activation function, and below it will be compared to that of the Exponential activation function.  

For the case of the Exponential (Poisson) activation function, the cumulant generating function is 
\begin{equation}
K^{\rm exp}_{\mu}(q_{\mu})=\exp (-c_{\mu})\left[\exp(q_{\mu})-1  \right].
\label{eq:Kexp}
\end{equation}
By plugging Eq.~\eqref{eq:Kexp} into Eq.~\eqref{exact}, the sum over visible indices factorizes and the expected value of the $s-$th order interactions between visible nodes with indices $\{k_1,k_2, \cdots k_s\}$ becomes
\begin{equation}
I^{\rm Exp}_s  =  \left \langle I^{\rm Exp}_{k_1,\cdots, k_s}\right \rangle = \int \prod_{k,\mu}dw_{k,\mu}  P(\{w_{k\mu}\}) P(\{c_{\mu}\})\; \sum_{\nu} \exp\left(-c_{\nu}\right) \prod_{j=1}^{s} \left( \exp w_{j\nu}-1\right) = M \gamma_1^s \langle  \exp\left(-c_{\mu}\right)\rangle 
\label{exp_exact}
\end{equation}
where 
\begin{equation}
\gamma_1 \equiv \langle \exp w \rangle-1\label{gamma1}.
\end{equation}
The first point to note here is that while for the case of the Linear activation function the expected values of the induced interactions depend only on the first moments of the distributions $p(w_{i\mu})$ and $p(c_{\mu})$, for the Exponential activation function, all the moments are important in determining $I^{\rm Exp}_{s}$.

Turning to the correlations between the interactions, consider two sets of nodes $\{k'_1,k'_2, \cdots k'_{s'}\}$ and $\{k_1,k_2, \cdots k_s\}$ with $s'\le s$ and $m$ nodes in common. As derived in Appendix \ref{app_expcorr}, the correlation between the induced interaction terms for each set can be written as
\begin{equation}
\left \langle I^{\rm Exp}_{k_1,\cdots, k_s} I^{\rm Exp}_{k'_1,\cdots, k'_{s'}} \right \rangle - \left \langle I^{\rm Exp}_{k_1,\cdots, k_s} \right \rangle \left \langle I^{\rm Exp}_{k'_1,\cdots, k'_{s'}} \right \rangle   = M
\gamma_1^{s+s'} \biggl[ \gamma_2^m \gamma_1^{-2m}-1\biggr] \langle \exp\left(-2c_{\mu}\right)\rangle
\label{exp_exact2}
\end{equation}
where 
\begin{equation}
\gamma_2 \equiv \left\langle \left(\exp w-1\right)^2\right\rangle.
\label{gamma2}
\end{equation}
Focusing on the case of $s=2$, we have that
\begin{equation*}
\left \langle {I^{\rm Exp}_{k_1,k_2}}^2 \right \rangle - \left \langle I^{\rm Exp}_{k_1,k_2} \right \rangle^2  = M \gamma^{4}_2 \left[\gamma^2_{2}\gamma^4_1-1\right] \langle \exp\left(-2c_{\mu}\right)\rangle
\end{equation*}
\begin{equation*}
\left \langle I^{\rm Exp}_{k_1,k_2} I^{\rm Exp}_{k_1,k_3} \right \rangle - \left \langle I^{\rm Exp}_{k_1,k_2} \right\rangle \left\langle I^{\rm Exp}_{k_1,k_3} \right \rangle = M \gamma^{4}_2 \left[\gamma_{2}\gamma^2_1-1\right] \langle \exp\left(-2c_{\mu}\right)\rangle, \; k_2\neq k_3
\end{equation*}
Note once more that, unlike in the case of the Linear activation function, the correlations now depend on all moments of $p(w_{i\mu})$ and $p(c_{\mu})$, even when considering only pairwise interactions.

Next, let us focus on the dispersion of the $s$th order interactions, that is their fluctuations around the expected values
\begin{equation*}
\Delta^{\rm Exp}_{s}\equiv \frac{\left \langle {I^{\rm Exp}_{k_1,\cdots, k_s}}^2 \right \rangle - \left \langle I^{\rm Exp}_{k_1,\cdots, k_{s}} \right \rangle^2}{\left \langle I^{\rm Exp}_{k_1,\cdots, k_{s}} \right \rangle^2} = M^{-1}_0
\left [(\gamma_2/\gamma^2_1)^{s}-1\right] ,
\end{equation*}
where 
$M_0 \equiv M \frac{\langle \exp(-c_{\mu}\rangle)^2}{\langle \exp(-2c_{\mu}) \rangle}$. The term multiplying $M$ in the above expression can be considered as a measure of the sparsity of the distribution of inputs: if $c_{\mu} = 0$ with a probability $p_0\ll 1$, then this term behaves as $p_0$. So, $M_0$ can be thought of as the effective number of hidden units receiving external input.

As $\gamma_1\to 0$, the expected values go to zero as $I^{\rm Exp}_s \sim \gamma^s_1$, while the corresponding fluctuation to mean ratio diverges as $\Delta^{\rm Exp}_s\sim \gamma^{-2s}_1$.
Furthermore, since $\gamma_2\ge \gamma^2_1$, we see that for $s>s'$, $\Delta^{\rm Exp}_s>\Delta^{\rm Exp}_{s'}$. This is true regardless of whether $I^{\rm Exp}_{s}$ is larger or smaller than $I^{\rm Exp}_{s'}$, which is determined by whether or not $\gamma_1>1$. Obviously, this means that the region of parameters (that define the distribution over the weights) for which $\Delta^{\rm Exp}_{s}>1$, that is where the fluctuation of the $s$-th order interactions are larger than their expected values, contains the region for which $\Delta^{\rm Exp}_{s'}>1$.\\  
In fact defining $a_s(M_0)\equiv(M_0+1)^{1/s}$, then
\begin{equation*}
\gamma_2 = \gamma^2_1 a_s(M_0),
\end{equation*}
defines the critical line on which $\Delta^{\rm Exp}_{s}=1$. When $\gamma_2 > \gamma^2_1 a_s(M_0)$, then the fluctuations are larger than the expected value, while for $\gamma_2 < \gamma^2_1  a_s(M_0)$ they are smaller. 

\begin{figure}[htbp]
\centering
\includegraphics[width=9cm]{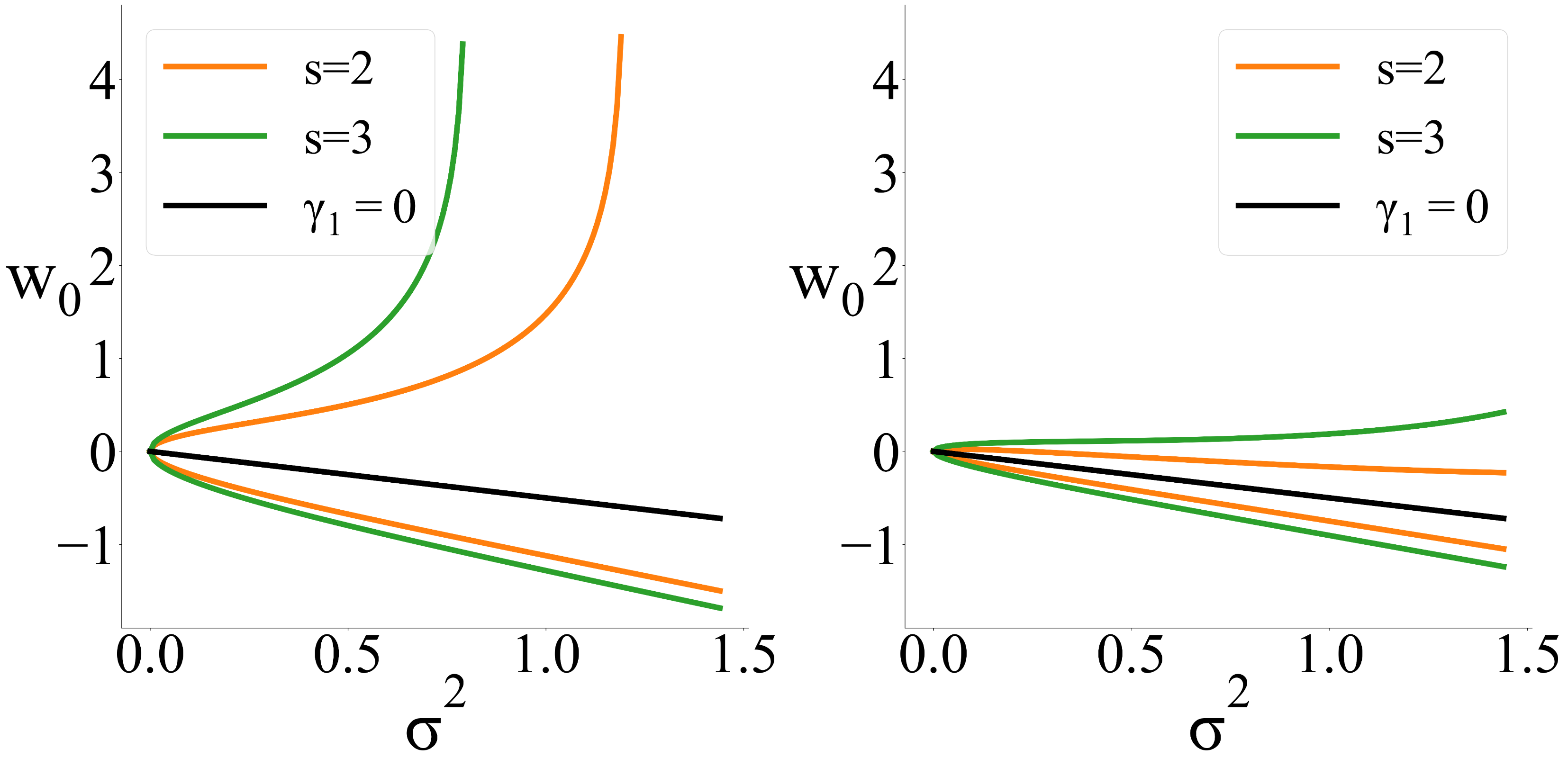}
\caption{Solutions of $\Delta^{\rm Exp}_{s}=1$ in the $(\sigma^2,w_0)$ plane for $M_0^{-1}=0.1$ (left) and $M_0^{-1}=0.002$ (right). Eq.~\eqref{delta_line} is plotted with a color corresponding to the order of interaction. The black line shows the divergence $\gamma_1=0$, where interaction fluctuations are infinitely larger than the expected value. }
\label{fig:exp_delta1}
\end{figure}

In the case of the Linear activation function, it is only the second order interactions that can be non-zero. All higher order interactions are zero and thus smaller than the pairwise interactions. The situation with the Exponential activation function is more interesting. Firstly,  Eq.~\eqref{exp_exact} shows that the expected value of the interaction term in the case of the Exponential activation function can increase exponentially with $s$ if $\gamma_1>1$. Let us consider weights with mean $w_0$ and variance $\sigma^2$. We then have 
\begin{equation}
\frac{I^{\rm Exp}_{s+1}}{I^{\rm Exp}_s}=\gamma_1
\label{transition_exp}
\end{equation}
Thus, $\gamma_1=1$ is where the expected values of all interactions are equal. For $\gamma_1>1$ they exponentially increase with $s$, while for $\gamma_1<1$ they exponentially decrease.\\The conclusions reached up to this point do not depend on the specific choice of distribution of $p(w_{i\mu})$ and $p(c_{\mu})$.
Let us now restrict ourselves to the case of Gaussian distributed weights. We then have $\gamma_1=\exp(w_0+\sigma^2/2)-1$ and $\gamma_2 = \exp(2w_0+2\sigma^2)-2\exp(w_0+\sigma^2/2)+1=\exp\left(\sigma^2\right)(\gamma_1+1)^2-2\gamma_1-1$. In this case, approaching the curve $w_0+\sigma^2/2=0$ (i.e. $\gamma_1=0$), the expected interactions $I^{\rm Exp}_{s}$ approach zero and $\Delta^{\rm Exp}_s$ diverges. Furthermore,
\begin{equation}
w_0+\sigma^2/2 = \log 2
\label{eq:exponential_transition}
\end{equation}
defines a critical line where the expected value of the $(s+1)$-th order interaction equals those of the $s$-th order interactions. On one side of this line, the former is larger than the latter. Note that the equation defining this line is independent of $s$, $M$ and the distribution of the inputs $c_{\mu}$.\\ Finally, the fluctuations become of the same order as the expected values, that is $\Delta^{\rm Exp}_s=1$, when 
\begin{equation}
    w_0^{\pm} = -\frac{\sigma^2}{2}- \log \Bigg[ 1\pm \sqrt{\left(\exp{ \sigma^2}-1\right)/(a_s(M_0)-1)}\Bigg]
    \label{delta_line}
\end{equation}
Increasing $\sigma$, $w^{+}_0$ increases until it diverges when $\exp\left(\sigma^2\right)=a_{s}(M_0)$. Increasing $\sigma$ further, $\Delta^{\rm Exp}_s$ exceeds one when $w_0>0$. $w^{-}_0$, however remains finite for all values of $\sigma$ and this is shown in Figure \ref{fig:exp_delta1}.
\\
Moving away from the divergence line $\gamma_1=0$, lower orders of interaction cross the lines defined by Eq.~\eqref{delta_line} and enter the $\Delta^{\rm Exp}_s<1$ region, before high order interactions do. In other words, 
interaction terms with increasing order $s$ have larger fluctuation-dominated regions. Furthermore, the size of these regions decreases linearly as the number of hidden nodes, $M$, increases.

To summarize, in this section we studied the expected value, fluctuations, and correlations of the interactions induced on visible units by hidden nodes whose activation is regulated by means of a Linear or Exponential function. In both cases, we derived analytical expressions for these quantities as a function of the distributions of the weights and fields acting on the hidden units. In particular, for the Exponential activation function we found that the expected value of a given interaction can increase with the order $s$. Furthermore, the first and second moment statistics of the interactions depend on the quantities $\gamma_1$ and $\gamma_2$ involving all moments of the distributions of the weights and fields acting on hidden units. We also found the condition under which the fluctuations in the interactions change from being smaller to larger than their expected values.

In the following section, we compute these expectations and fluctuations as second-order expansions of Eq.~\eqref{exact} in the weights. This will allow us, at least for small weight fluctuations, to study the statistics of the interactions for other activation functions. We apply this approach to calculate expectations and fluctuations for the Step and ReLU activation functions and also compare the results with Linear and, in particular, Exponential activations. 

\subsection{Interactions for small fluctuations}

An approximation for the expected values and fluctuations can be derived by considering $w_{i\mu} = w_{0}+\delta w_{i\mu}$, where $\delta w_{i\mu}$ are fluctuations around $w_{0}$ and expanding the expression in Eq.~\eqref{exact} around $w_{0}$. This corresponds to expanding the cumulant generating function $K_{\mu}$ around $(s-p)w_0$, in each term in the sum over $p$ of Eq.~\eqref{exact}\footnote{To be more precise, the expansion is valid when $(s-p)\delta w_{i\mu}$ is small. For this to hold it is sufficient that $s\delta w_{i\mu}$ is small.}. The resulting interaction term will take the form
\begin{equation}
    I_{k_{1},\cdots,k_{s}} \equiv I_{0}^{(s)} + \delta I_{k_{1},\cdots,k_{s}}
    \label{inter_terms}
\end{equation}
where
\begin{equation}
    I_{0}^{(s)} \equiv \sum_{\mu}\sum_{p=0}^{s-1}(-1)^{p} \binom{s}{s-p}  K_{\mu}\left((s-p)w_{0}\right)
    \label{i0}
\end{equation}
is a deterministic term that only depends on the expected value $w_{0}=\langle w_{i\mu}\rangle$. Once an activation function is chosen, this term is completely determined by the order of interaction, the number of hidden nodes and $w_{0}$. The fluctuating term $\delta I_{k_{1},\cdots,k_{s}}$ can be shown to be (see Appendix \ref{appB}): 
\begin{equation}
    \delta I_{k_{1},\cdots,k_{s}}=\sum_{\mu}\alpha_{s\mu}(w_0) \sum_{i}^{s} \delta w_{k_{i}\mu} + \sum_{\mu}\beta_{s\mu}(w_0) \sum_{i}^{s} \delta w_{k_{i}\mu}^{2} + \sum_{\mu}\eta_{s\mu}(w_0) \sum_{i<j}^{s}  \delta w_{k_{i}\mu}\delta w_{k_{j}\mu}
    \label{fluct_comp}
\end{equation}
where 
\begin{align}
&\alpha_{s\mu}(w_{0}) \equiv \sum_{p=0}^{s-1}(-1)^{p}\binom{s-1}{s-1-p} K_{\mu}^{'}\left((s-p)w_{0}\right),
\label{alpha}\\
&\beta_{s\mu}(w_{0}) \equiv \sum_{p=0}^{s-1}\frac{(-1)^{p}}{2}\binom{s-1}{s-1-p} K_{\mu}^{''}\left((s-p)w_{0}\right),
    \label{beta}\\
&\eta_{s\mu}(w_0) = \sum_{p=0}^{s-2}(-1)^{p}\binom{s-2}{s-2-p} K_{\mu}^{''}\left((s-p)w_{0}\right)
    \label{eta}
\end{align}
The last term in Eq.~\eqref{fluct_comp} takes the form of Hebbian learning which has been noted for the pairwise interactions of the Linear activation function e.g. in \cite{LEONELLI2021314}, and for small weights in \cite{10.1162/neco_a_01420}.
The small parameters expansion of \cite{10.1162/neco_a_01420} for the pairwise interactions is a specific case of Eq.~\eqref{fluct_comp} when $w_{0}=0$. Given that $\alpha_{\mu}^{(s)}(0)=0$ and $\beta_{\mu}^{(s)}(0)=0$, $\forall s=2,...,N$
\begin{equation}
    I_{k_{1},k_{2}}\biggl|_{w_{0}=0} =\sum_{\mu}\eta_{2\mu}(0)\sum_{i<j}^{2}  \delta w_{k_{i}\mu}\delta w_{k_{j}\mu}=\sum_{\mu}k^{(2)}_{\mu}\delta w_{k_{1}\mu}\delta w_{k_{2}\mu}.
    \label{pair_small}
\end{equation}
where $k_{\mu}^{(2)}$ is the second cumulant of $\rho$ defined in Eq.~2.5 of \cite{10.1162/neco_a_01420}. This expression generalizes to the leading term for higher order interactions $I_{k_{1},\cdots,k_{s}}\biggl|_{w_{0}=0}\simeq \sum_{\mu}k_{\mu}^{(s)}w_{k_1\mu}\cdots w_{k_{s}\mu}$. Eq.~\eqref{inter_terms} thus generalizes the small parameter expansion to when $w_0\neq0$ and to higher order interactions. The interesting point to note here is that besides its dependence on the activation function, the strength of the Hopfield term present in the $s$-th order interactions depends on $s$ via $\eta_{s\mu}$. In principle, then the Hopfield term can have a larger or smaller effect on the different orders of interactions $s$ depending on the second derivative of $K_{\mu}$. For the case of Linear activation function $\eta^{\rm Lin}_{s\mu}=1$. Instead, for the Exponential activation function $\eta^{\rm Exp}_{s\mu}=\exp\left(-c_\mu+2w_0\right) (\exp w_0-1)^{s-2}$.

\subsection{Expectations over an ensemble of random RBMs}

We can use Eq.~\eqref{fluct_comp} to compute the statistical properties of the $s$-th order interaction within the small weight fluctuations expansion for an ensemble of RBMs. We consider an ensemble with weights such that
\begin{eqnarray*}
    \langle w_{i\mu} \rangle &=& w_0,\\
    \langle \delta w_{i\mu} \rangle &=& 0 \qquad \text{and} \qquad  
    \langle \delta w_{i\mu} \delta w_{j\nu} \rangle  =  \sigma^2 \delta_{ij} \delta_{\mu\nu}
\end{eqnarray*}
where $\delta w_{i\mu}=w_{i\mu}-w_0$, $\sigma = g/\sqrt{M}$ and $g$ is a positive constant.\\ Within these assumptions, we have that
\begin{equation*}
   \delta I_s \equiv \left\langle \delta I_{k_{1},\cdots,k_{s}}\right\rangle = sg^{2}\widehat{\beta^{(s)}}
\end{equation*}
where $\widehat{\beta^{(s)}} = \frac{1}{M} \sum_{\mu}\beta_{\mu}^{(s)}$ which means 
\begin{equation}
    I_s= I_{0}^{(s)} + sg^2\widehat{\beta^{(s)}}.
    \label{expected}
\end{equation}
Under the same assumptions, the covariances can be computed, and the second moment takes the form
$\left\langle {\delta I_{k_{1},\cdots,k_{s}}}^2 \right\rangle
    =  sg^2 {\widehat{{\alpha^{(s)}}^{2}}} + \mathcal{O}(\delta w^3)$.\\
The covariances can be computed between arbitrary pairs sharing $q$ visible nodes for the Gaussian ensemble (Appendix \ref{appC}), and the variance is
\begin{eqnarray}
    \left\langle \delta {I_{k_{1},\cdots,k_{s}}}^{2}  \right\rangle-\left\langle \delta {I_{k_{1},\cdots,k_{s}}}  \right\rangle^2 & = & s\sigma^{2} \sum_{\mu} \left[  \alpha_{\mu}^{2} + 2\sigma^{2}\beta_{\mu}^{2} + \frac{s-1}{2}\sigma^{2}\eta_{\mu}^{2}\right].
    \label{eq:variance_fluct}
\end{eqnarray}
For a large number of hidden nodes $(M\to \infty)$  we get 
    $\left\langle {\delta I_{k_{1},\cdots,k_{s}}}^{2}  \right\rangle \simeq sg^{2} \left[{\widehat{{\alpha^{(s)}}^{2}}} + sg^{2}{\widehat{{\beta^{(s)}}^{2}}}\right].$\\
Then, the total second moment of the interaction can be computed as 
\begin{equation}
    I_{s,2}\equiv \left\langle {I_{k_{1},\cdots,k_{s}}}^2 \right\rangle = \left\langle \left(I_0^{(s)} + \delta I_{k_1,\cdots, k_s}\right)^2\right\rangle
    \label{moment2}
\end{equation}
Equations~\eqref{i0},~\eqref{expected} and~\eqref{eq:variance_fluct} are general for any activation function in the approximation of small weight variability. As reported in section \ref{3.1}, the exact moments, even when the weight variability is not small, can be computed for an ensemble of RBMs with Exponential and Linear activation functions, yielding Eqs.~\eqref{exp_exact} and~\eqref{exp_exact2}.

In the next section, we compare the analytical expressions for the first and second moments of the $s$-th order interactions with their approximate values computed above. Because we are ultimately interested in finding out how the expected/ensemble averages reflect the statistical properties of a single RBM, we also compare the empirical averages
\begin{equation}
    \overline{ I_s^n} \equiv \binom{N}{s}^{-1}\sum_{k_{1}<...<k_{s}}^{N} \left[ {I_{k_{1},\cdots,k_{s}}}\right]^n
    \label{empirical}
\end{equation}
where the sum goes over all visible units in a {\em single} RBM chosen from the ensemble with Gaussian weights.

\subsection{Interactions state space}
\begin{figure}[htbp]
\centering
\includegraphics[width=\textwidth]{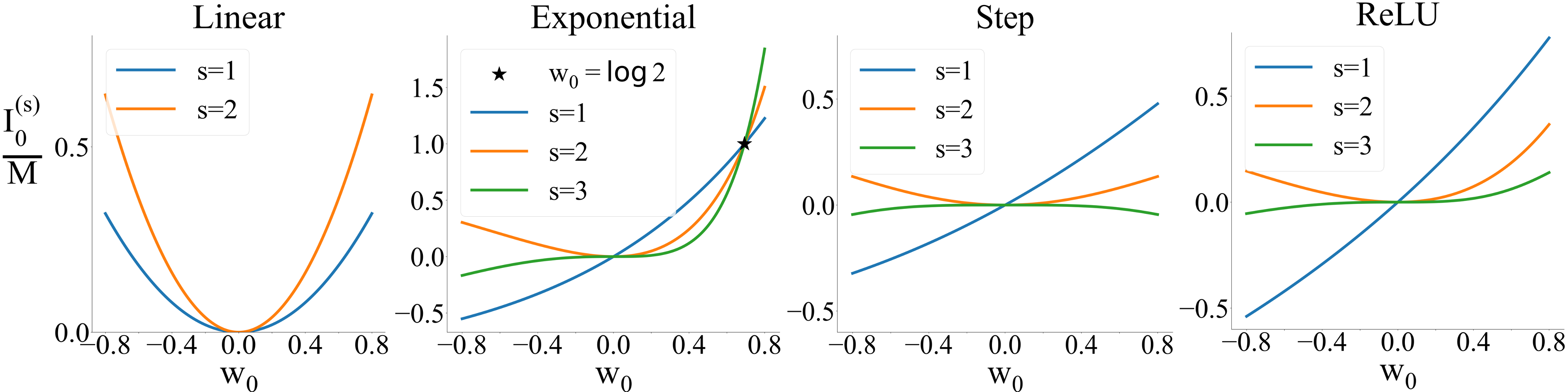}
\caption{$I_0^{(s)}/M$ versus $w_0$ from Eq.~\eqref{i0}, for $s=1,2$ for the Linear activation function and $s=1,2,3$ for Exponential, Step and ReLU. Interactions of higher orders are also present for all the activation functions except for Linear. $I_0^{(s)}/M$ with $s>3$ are smaller than $I_0^{(3)}/M$ and are not shown for visualization purposes. The star indicates the transition point for the Exponential function. The RBM parameters are $b_i=0$ $\forall i$, $c_{\mu}=0$, $\forall \mu$.}
\label{fig:I_0}
\end{figure}

\begin{figure}[htbp]
\centering
\includegraphics[width=16 cm]{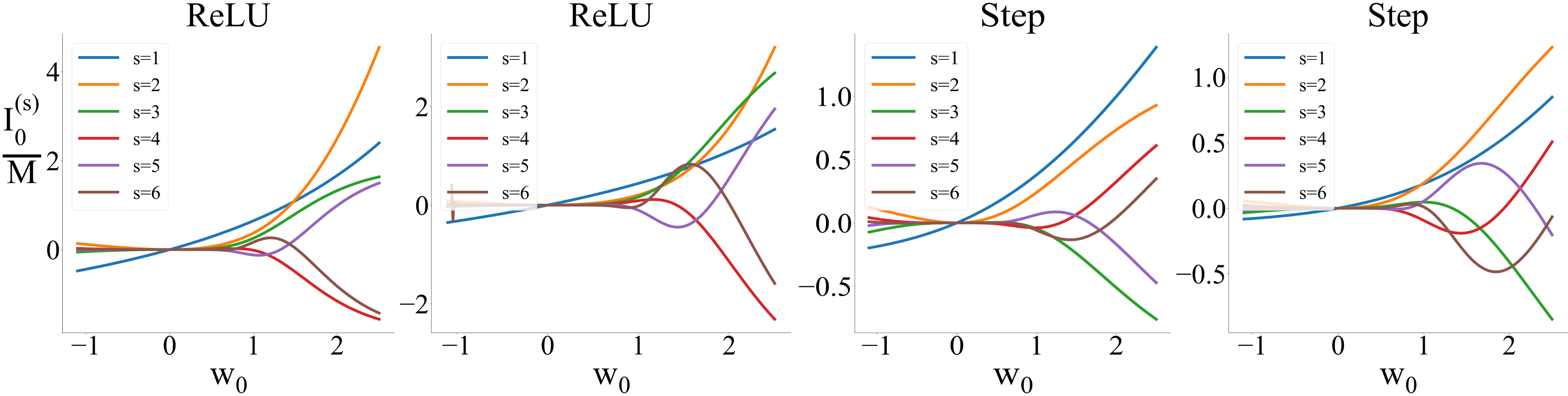}
\caption{$I^{(s)}_0/M$ versus $w_0$ for $c=1$ (first and third panels) and $c=2$ (second and fourth panels) for ReLU and Step.}
\label{fig:cdep}
\end{figure}

\begin{figure}[htbp]
\centering
\includegraphics[width=\textwidth]{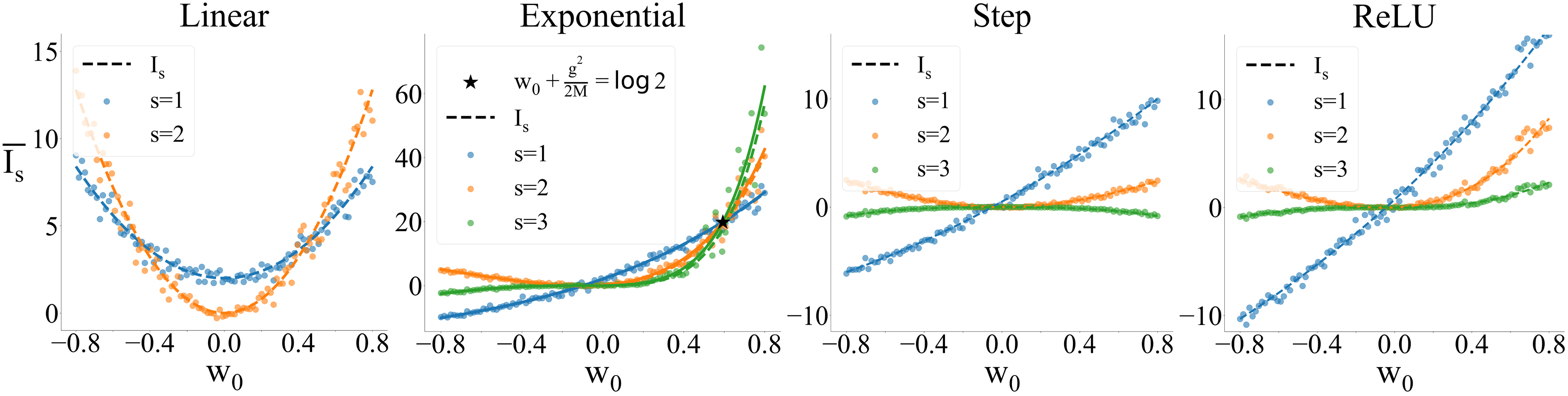}
\caption{$\overline{I_s}$ from Eq.~\eqref{empirical} and $I_s$ from Eq.~\eqref{expected} (dashed line)  versus $w_0$ for $g=2$. The solid line for the Exponential activation shows Eq.~\eqref{exp_exact}. Parameters are $b_i=0$, $c_{\mu}=0$ $\forall i,\mu$, $N=8$ and $M=20$.}
\label{fig:I_avg_w0}
\end{figure}

We start by comparing empirical moments of a single RBM interactions with ensemble averages in Eqs.~\eqref{expected} and~\eqref{eq:variance_fluct}. In Figure \ref{fig:I_0}, we first plot $I_0^{(s)}/M$ for different activation functions for $c_{\mu}=0$. We have also plotted the expected value of $I_i(w_0)=b_{i}+\sum_{\mu} K(w_{i\mu})$ but only for $b_i=0$ as $b_{i}$ only appears as an additive term in $I_i$ and does not have any other effect on the interactions. 
The transition point in Eq.~\eqref{eq:exponential_transition} for $\sigma =0$, that is $w_0 = \log 2$, is where  $I_0^{(s+1)}/I_0^{(s)}=1$. This is also shown in the plot for the Exponential activation function. For $w_0>\log 2$, the deterministic part in the expansion of the interactions of order $s+1$ becomes larger than the ones of order $s$, for all $s$.\\
For ReLU, Step and Exponential activation functions, $I_0^{(1)}$ is an increasing function of $w_0$ and $I_0^{(2)}$ an increasing function of $|w_0|$. The rates of increase in Exponential are much faster, while ReLU and Step have very similar rates of increase. While $I_0^{(3)}$ is always negative for Step, it takes the same sign as $w_0$ for ReLU. For the Exponential activation function, the effect of turning on the external input to the hidden nodes $c_{\mu}$ is a multiplication of all interactions by the same number $\sum_{\mu} \exp\left(-c_{\mu}\right)$. The relative magnitude of $I^{\rm Exp}_s$ thus does not change if $c_{\mu}\neq 0$. Fig.~\ref{fig:cdep} shows a different situation
for ReLU and Step, where the input to the hidden units changes the interaction structure in a more complicated way.\\
Turning to the empirical values, Fig.~\ref{fig:I_avg_w0} shows how the values of $\overline{I}_s$ defined in Eq.~\eqref{empirical} ($n=1$) compare with the results of the expansion for a range of $w_0$ and $g=2$. In the case of Exponential, the analytical expressions from Eq.~\eqref{exp_exact} are also shown.

\begin{figure}[htbp]
\centering
\includegraphics[width=16.1 cm]{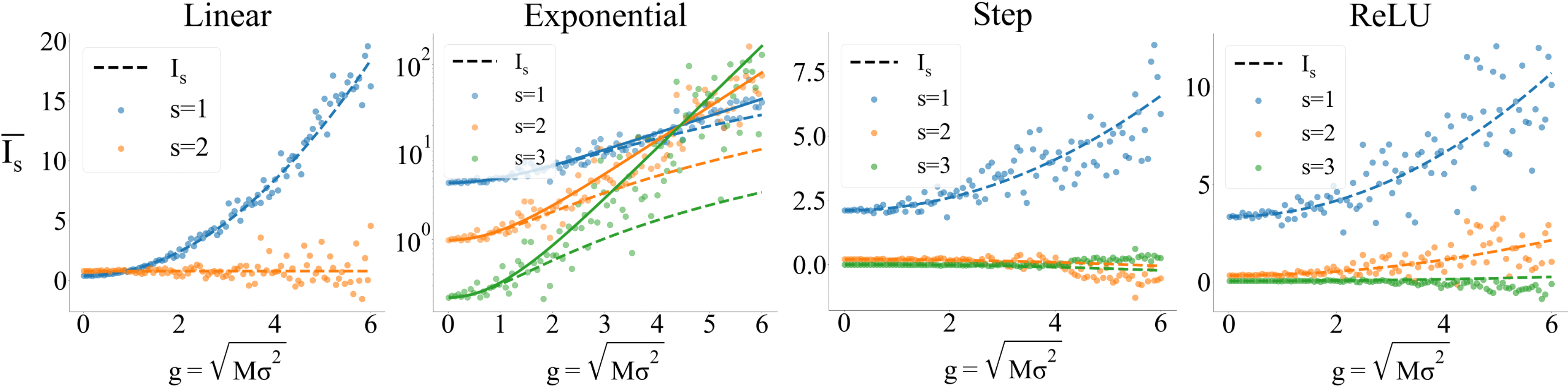}
\caption{$\overline{I_s}$ from Eq.~\eqref{empirical} and $I_s$ from Eq.~\eqref{expected} (dashed line) versus $g$ for $w_0=0.2$. The solid line for the Exponential activation shows Eq.~\eqref{exp_exact}. The RBM parameters are $b_i=0$, $c_{\mu}=0$ $\forall i,\mu$, $N=8$ and $M=20$.}
\label{fig:I_avg_g}
\end{figure}

\begin{figure}[htbp]
\centering
\includegraphics[width=\textwidth]{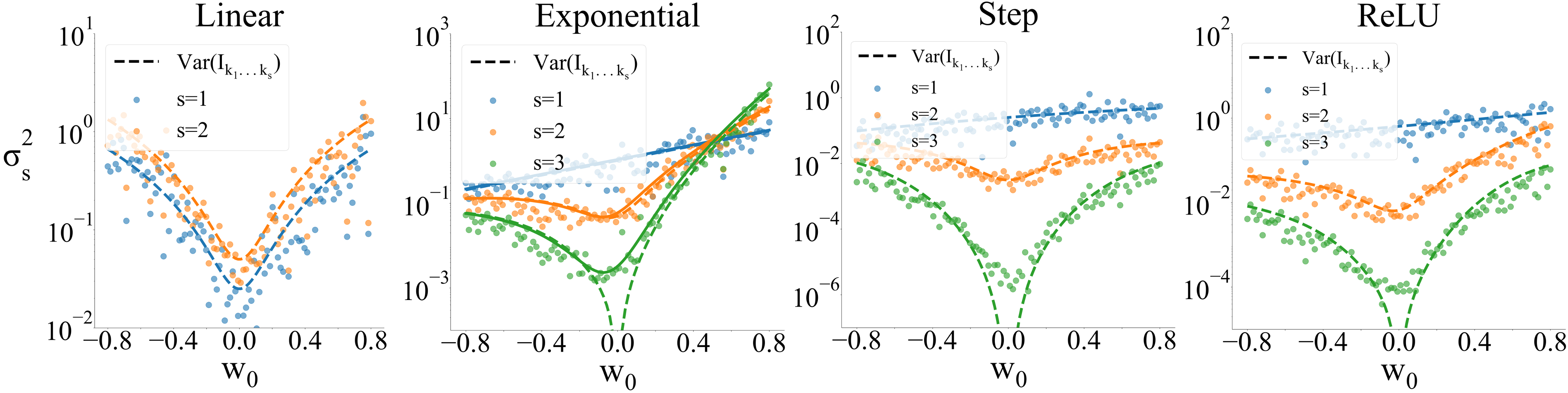}
\caption{$\sigma_s^2$ from Eq.~\eqref{emp_variance} and ${\rm Var}(I_{k_{1},\cdots,k_{s}})$ from Eq.~\eqref{variance_I} (dashed line) versus $w_0$ for $g=1$. The solid line for the Exponential activation shows Eq.~\eqref{exp_exact2}. The RBM parameters are $b_i=0$ $\forall i$, $c_{\mu}=0$ $\forall \mu$, $N=8$ and $M=20$.}
\label{fig:I_var_w0}
\end{figure}

\begin{figure}[htbp]
\centering
\includegraphics[width=\textwidth]{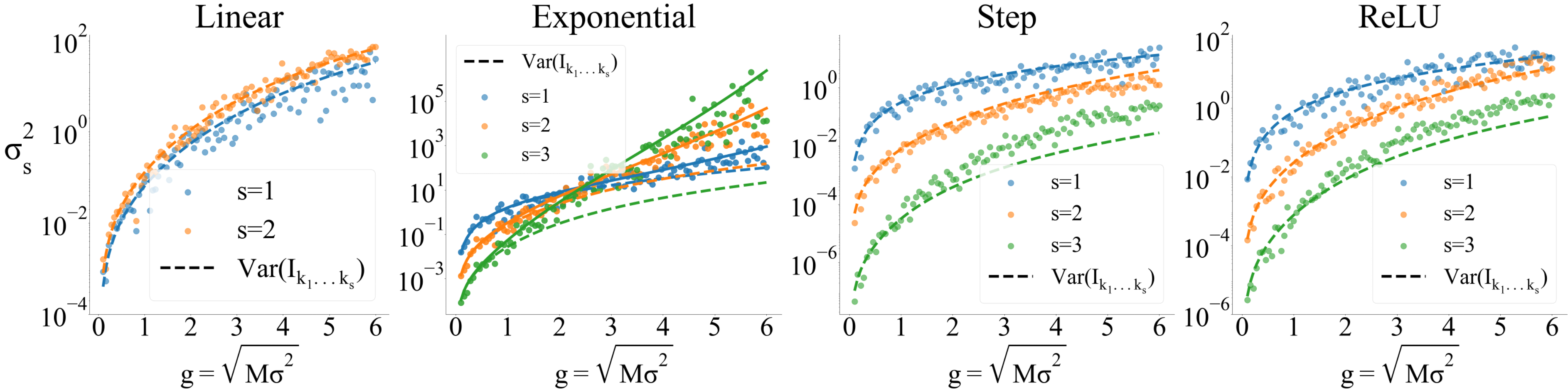}
\caption{$\sigma_s^2$ from Eq.~\eqref{emp_variance} and ${\rm Var}(I_{k_{1},\cdots,k_{s}})$ from Eq.~\eqref{variance_I} (dashed line) versus $g$ for $w_0=0.2$. The solid line for the Exponential activation shows Eq.~\eqref{exp_exact2}. The RBM parameters are $b_i=0$ $\forall i$, $c_{\mu}=0$ $\forall \mu$, $N=8$ and $M=20$.}
\label{fig:I_var_g}
\end{figure}

The introduction of small weight variability slightly changes the curves of $\overline{I}_s$ compared to the deterministic term $I^{(s)}_0$, for each activation function.
As shown in the figure, for $g=2$ there is a very good agreement between the empirical and theoretical values.\\
By showing the same quantities for increasing values of $g$ and $w_0=0.2$, Fig.~\ref{fig:I_avg_g} indicates how ensemble and empirical averages depart from each other. 
In general, for a wide range of $g$ there is a good agreement between the two, and as expected, deviations occur for large $g$. The deviations occur at lower $g$ for the Exponential activation function and are more significant.\\
In this case, the analytical values $ I^{\rm Exp}_s$ from Eq.~\eqref{exp_exact} show a good agreement with the empirical means even for large $g$, where the $\gamma_1=1$ transition takes place. The disagreement between the empirical values and second-order expansion also takes place where the latter deviates from the analytical expressions. In fact, the $\gamma_1=1$ transition is not captured by the expansion.\\
Figs.~\ref{fig:I_var_w0} and ~\ref{fig:I_var_g} show a similar set of results as in Figs.~\ref{fig:I_avg_w0} and ~\ref{fig:I_avg_g} but for the variance of the interactions and $g=2$.

As in the previous cases, analytical expressions are compared with empirical averages.\\
This time, the variance over the ensemble
\begin{equation}
    {\rm Var}(I_{k_{1},\cdots,k_{s}})\equiv I_{s,2}-{I_s}^2= \left\langle {I_{k_1,\cdots,k_s}}^2\right\rangle-\left\langle I_{k_1,\cdots,k_s}\right\rangle^2
    \label{variance_I}
\end{equation} 
is compared with the empirical variance 
\begin{equation}
    \sigma_s^2 \equiv \overline{I_s^2}-{\overline{I_s}}^2
    \label{emp_variance}
\end{equation}
In this case, also, for Exponential activation, there is a point in which the variance increases with the order of interaction $s$.

\begin{figure}[htbp]
\centering
\includegraphics[width=\textwidth]{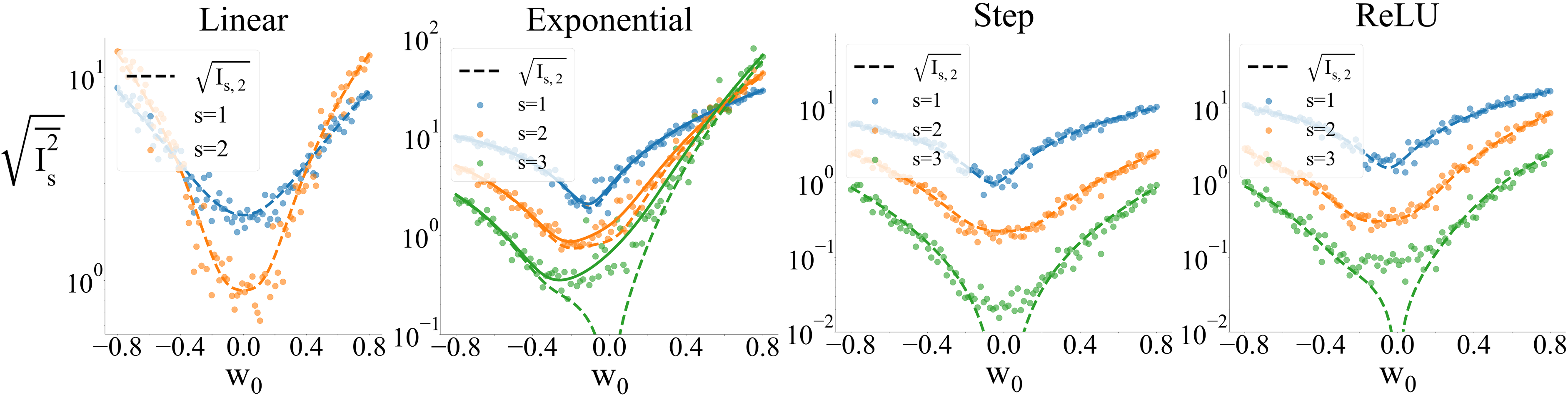}
\caption{Square root of $\overline{I_s^2}$ from Eq.~\eqref{empirical} ($n=2$) and square root of $I_{s,2}$ from Eq.~\eqref{moment2} (dashed line) versus $w_0$ for $g=1$. The solid line for the Exponential activation shows the first term in Eq.~\eqref{exp_exact2}. The RBM parameters are $b_i=0$ $\forall i$, $c_{\mu}=0$ $\forall \mu$, $N=8$ and $M=20$.}
\label{fig:I_mom2_w0}
\end{figure}

This is well captured by the theoretical expressions and shown both as a function of $w_0$ (Fig. \ref{fig:I_var_w0}) and as a function of $g$ (Fig. \ref{fig:I_var_g}). The same behavior is not seen for other activation functions and the variance always decreases with the order of interaction $s$.\\
Similar results are shown in Figs.~\ref{fig:I_mom2_w0} and ~\ref{fig:I_mom2_g} for the second moments both as a function of $w_0$ at $g=1$ and as a function of $g$ for $w_0=0.2$ (Figure~\ref{fig:I_mom2_g}).

\begin{figure}[htbp]
\centering
\includegraphics[width=5cm]{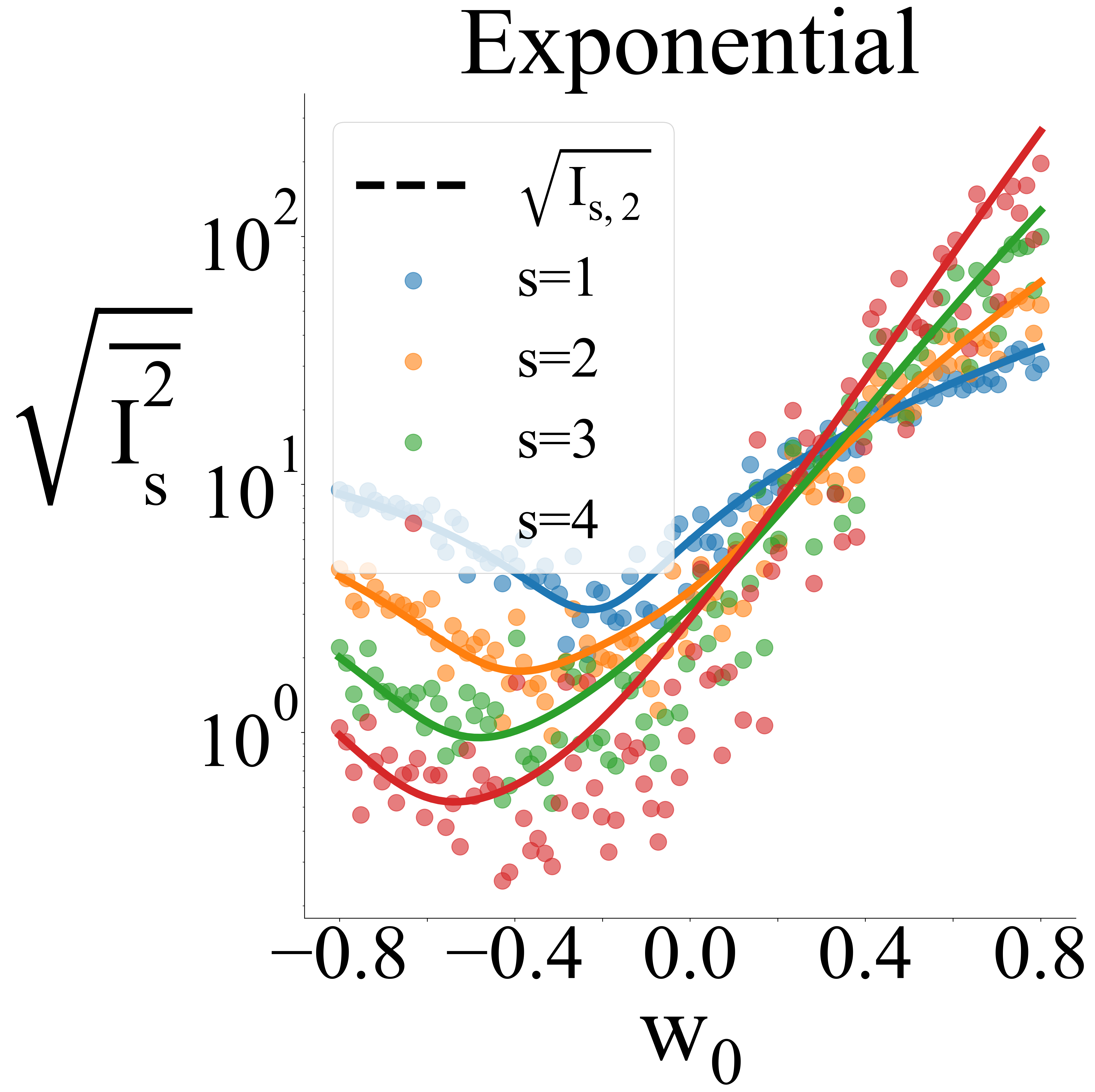}
\caption{RBMs with Exponential activation have a regime where different orders of interaction are of the same magnitude for small $w_0$. Eq.~\eqref{moment2} is computed for $g=2.85$, $N=8$ and $M=20$.}
\label{fig:I_mom2_w0_0}
\end{figure}

The square root of the interaction second moments gives an estimate of the average magnitude of the interaction terms. Then, the figures show how lower order interactions are larger in magnitude compared to high order interactions in most of the parameter space and for all the activation functions, except for Exponential. In this case, Fig.~\ref{fig:I_mom2_w0} shows that the second moments cross each other at a critical point similar to that of the expected values of the interactions, and so does the average interaction magnitude. This can be understood by noting that for large $M$, setting $s=s'$ in Eq.~\eqref{exp_exact2} becomes
\begin{equation*}
  I^{\rm Exp}_{s,2} \equiv \left\langle {I^{\rm Exp}_{k_{1},\cdots,k_{s}}}^{2}\right\rangle \sim M^{2}\gamma_1^{2s}
\end{equation*}
which results in the same transition point $\gamma_1=1$ for the second moments $I^{\rm Exp}_{s,2}$ as Eq.~\eqref{transition_exp} resulted for $I^{\rm Exp}_{s}$.\\ By increasing the value of $g$, the transition moves in the $(w_0,\sigma^2)$ plane so that the second moment of high-order interactions can be of the same order of magnitude as the one of lower orders, even for values of $w_0$ far from $w_0=\log 2$. This is shown in Fig. \ref{fig:I_mom2_w0_0} for the first four interaction second moments.
\\
In Figs.~\ref{fig:I_var_w0} and \ref{fig:I_mom2_w0}, small discrepancies between empirical and ensemble averages for $s>2$ for $w_0\sim 0$ are due to the fact that for an interaction of order $s$, the leading order in the expansion for $w_{i\mu}\sim 0$ is $s$ \cite{10.1162/neco_a_01420}, while the analytical expansion is up to second order. This also plays a role in the earlier departure of the theoretical line from the empirical values for $s=3$, compared to $s=1,2$, in Fig.~\ref{fig:I_var_g}.

\section{Learning decaying and non-decaying interaction models}
\label{sec4}
\begin{figure}[htbp]
\centering
\includegraphics[width=5cm]{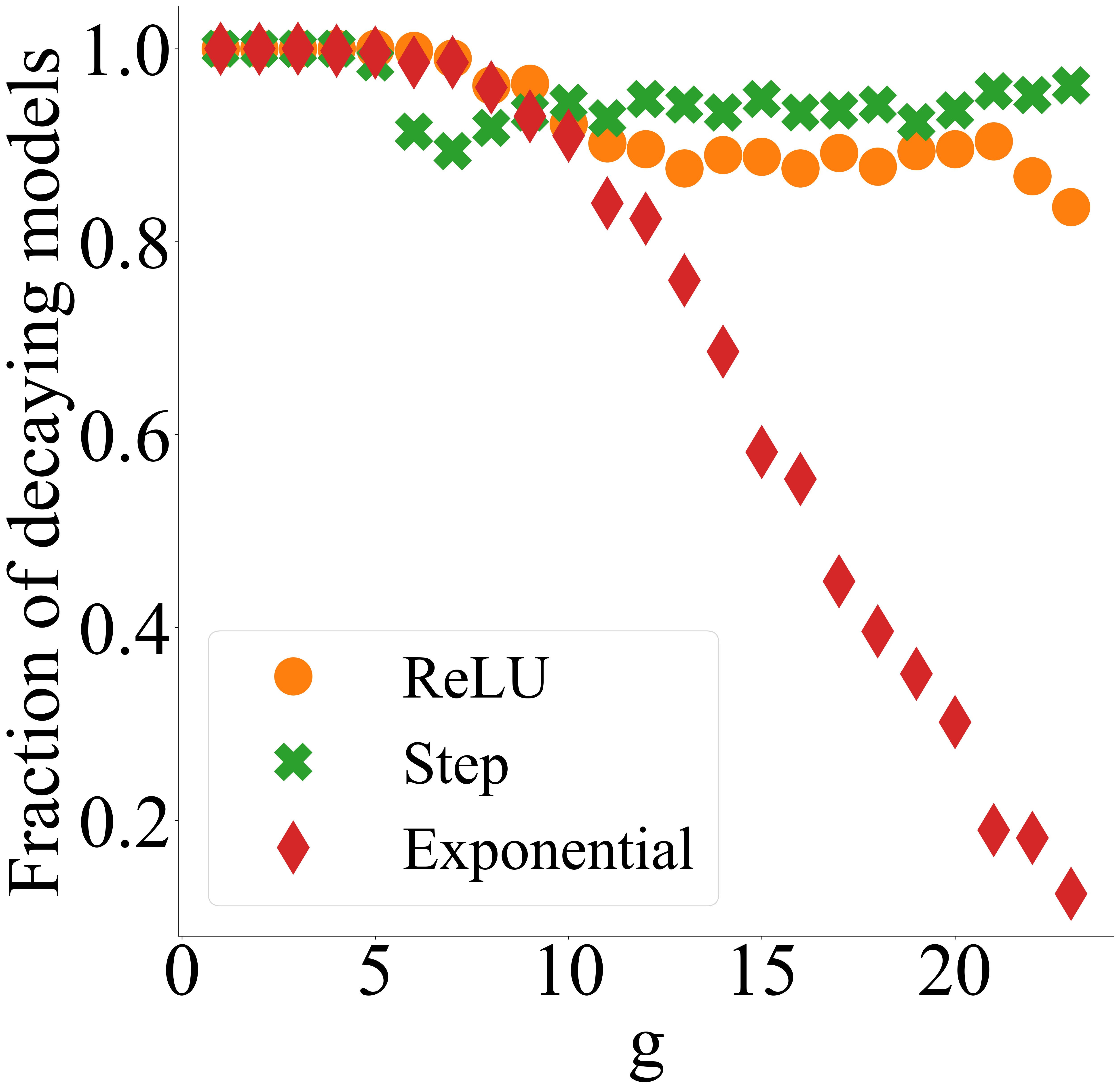}
\caption{Fraction of decaying interaction models for Exponential, ReLU and Step activation functions. A set of $500$ RBMs is generated with $N=5$, $M=500$ and Gaussian weights ($w_{0}=0$) and mapped into an interaction model for different values of $g$. Numerical precision is set to $10^{-4}$.}
\label{fig:fraction}
\end{figure}

\begin{figure}[htbp]
\centering
\includegraphics[width=8.5cm]{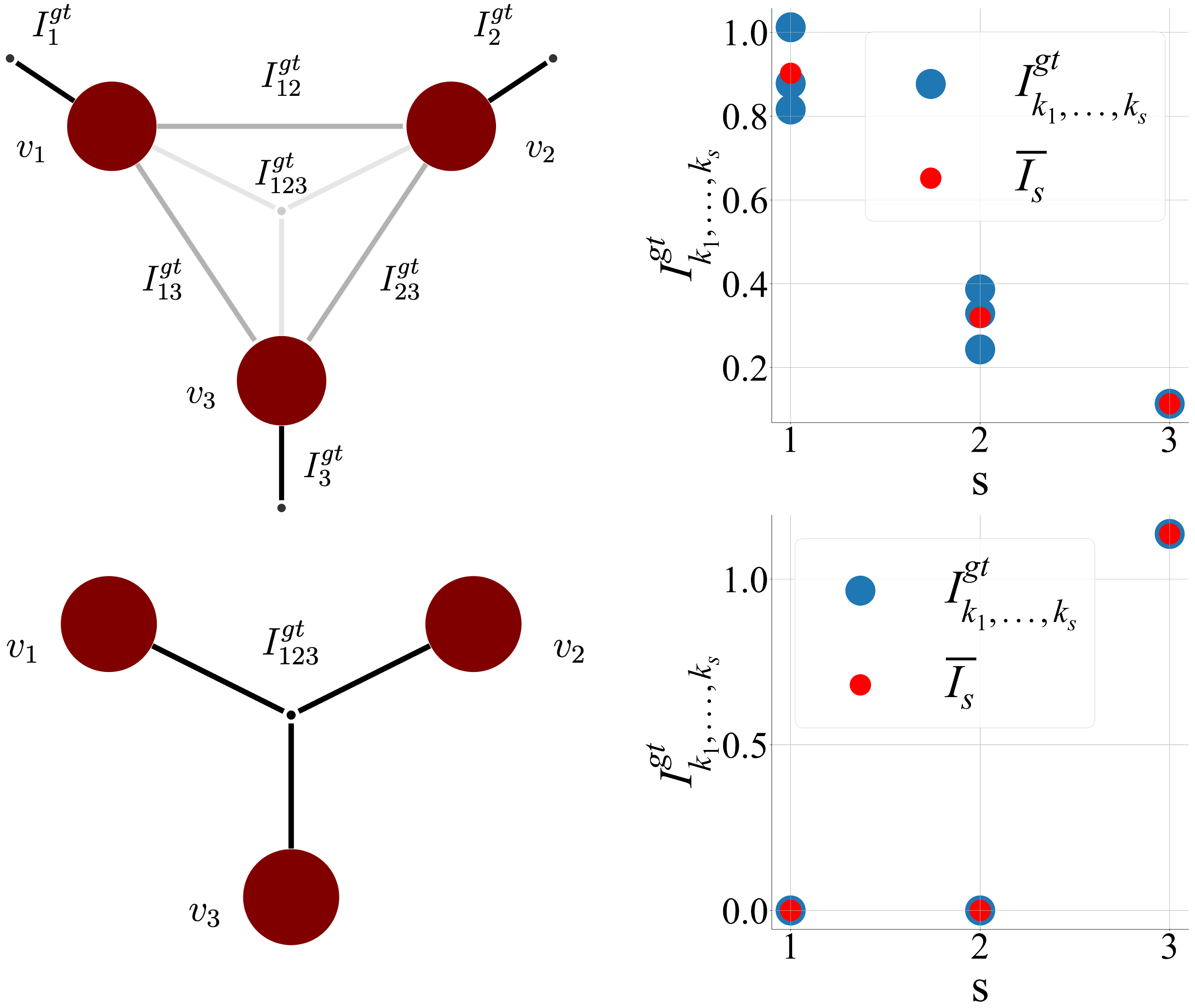}
\caption{\textbf{Decaying and non decaying ground truth lattice gas models}. Ground truth lattice gas models with $N=3$ and interactions $I^{gt}_{k_1,...,k_s}\sim \mathcal{N}(I^{(s)}_{gt}, I^{(s)}_{gt}/5)$. For the decaying interaction model in Eq.\eqref{dec_gt} {\red (upper left)}, $I^{(1)}_{gt}=0.9$, $I^{(2)}_{gt}=0.3$ and $I^{(3)}_{gt}=0.1$. For the non decaying interaction model in Eq.\eqref{dec_gt} {\red (lower left)}, the interactions are 3-body, $I^{(3)}_{gt}=1$. {\red Edges in the networks represent interaction terms with magnitude proportional to color shade. Interaction terms are shown versus $s$ on the right panels. It can be seen how interactions "decay" with $s$ for the decaying model (upper right) and don't "decay" for the non-decaying model (lower right).}}
\label{fig:NR_1}
\end{figure}

\begin{figure}[htbp]
\centering
\includegraphics[width=\textwidth]{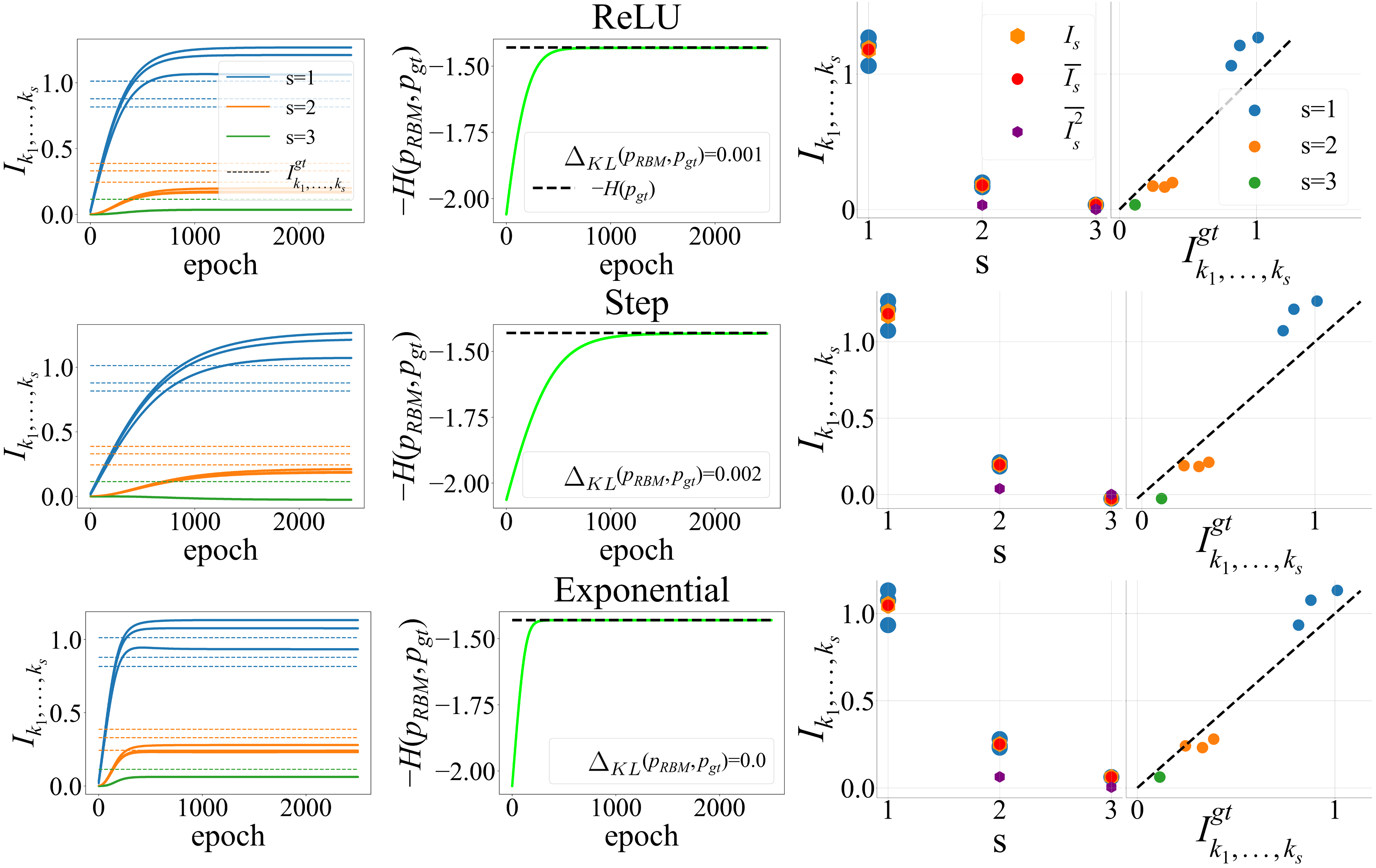}
\caption{Learning a decaying interaction model. A RBM with $N=3$ and $M=4$, initialized with zero-mean Gaussian weights ($\sigma=0.01$), is trained to match Eq.\eqref{dec_gt} for different activation functions. The model is trained for 2500 epochs with a learning rate of 0.001. The first panel in each row shows the trajectory of the interactions mapped from the RBM, compared with the ground-truth interactions (dashed lines). The second panel in each row shows the cross-entropy trajectory, where the target is the ground truth entropy (dashed line). $\Delta_{KL}(p_{RBM},p_{gt})$ is reported in the legend for the RBM at the end of training.  The third panel in each row shows the interactions mapped from the trained RBM, with empirical and expected moments. The fourth panel shows the comparison between ground truth and trained interactions.}
\label{fig:NR_2}
\end{figure}

In the previous sections we derived theoretical results for the statistics of the interactions between visible nodes in ensembles of RBMs and compared the analytical expression with empirical averages. We did this for different hidden node activation functions either exactly or via an approximation, and compared the results with empirical values estimated from a single RBM.\\
What can we learn from this about the role that the activation functions play in the ability of RBMs to learn different distributions? One implication could be that the choice of activation function impacts the proficiency of an RBM in learning distributions with strong higher-order interactions. It is only in the case of Exponential that we see a transition point where expected higher-order interactions and their corresponding variances become larger than the lower-order ones. Consequently, in this case one expects that near the transition point, a large fraction of the RBMs in the ensemble have larger higher-order interactions than the lower-order ones. On the contrary, such RBMs are unlikely to occur in the ensemble.\\ To be more precise, let us define \textit{decaying} interaction models as those where
\begin{equation}
\frac{\overline{{I_{k_{1},\cdots,k_{s+1}}}^2}}{\overline{{I_{k_{1},\cdots,k_{s}}}^2}} <1 \qquad \forall s;
\label{decaying}
\end{equation} 
a \textit{non-decaying} interaction model is one for which the above condition is not true.\\
As shown in the previous sections, when the weights are small, Eq.\eqref{inter_terms} can be expanded around $w_0\sim 0$ giving $I_{k_{1},\cdots,k_{s}} \simeq \sum_{\mu}k_{\mu}^{(s)}w_{k_1\mu}\cdots w_{k_{s}\mu}$. Plugging this expression in Eq. \eqref{decaying}
\begin{equation}
\frac{\overline{{I_{k_{1},\cdots,k_{s+1}}}^2}}{\overline{{I_{k_{1},\cdots,k_{s}}}^2}}\approx \frac{\left\langle {I_{k_{1},\cdots,k_{s+1}}}^2\right\rangle}{\left\langle {I_{k_{1},\cdots,k_{s}}}^2\right\rangle} \propto \frac{\prod_{i}^{s+1}\langle {w_{k_{i}\mu}}^2\rangle}{\prod_{i}^{s}\langle {w_{k_{i}\mu}}^2\rangle}=g^2/M.
\end{equation}
For sufficiently small $g$ and $w_0$, then, for all activation functions the models are decaying. Increasing $g$, in the case of the Exponential activation function, one, however, reaches the transition point shown e.g. in Figure \ref{fig:I_mom2_w0}. Beyond this point, both the expected value and the variance of the higher order interactions are larger than those of the lower order ones. Thus, one enters a regime where the models are non-decaying. This is shown in Figure~\ref{fig:fraction}, 
where we counted the number of RBMs with decaying interactions in the random ensembles.
Figure \ref{fig:fraction} shows that the fraction of decaying models is close to one for small $g$, as predicted by the theoretical analysis, for Exponential, Step and ReLU activation functions. This quantity smoothly drops around $g \sim 10$ to enter the non-decaying regime for the Exponential activation, while it remains large for the other functions. This, in turn, implies that it would be easier for RBMs with Exponential activation to learn non-decaying interaction models than it is for the other activation functions, at least in a region of the parameter space.\\
Although training RBMs on data often involve many choices and perks, the argument above leads to the hypothesis that training RBMs on datasets with decaying interactions should generally lead to a trained RBM with decaying interactions, as those are a priori more abundant. On the other hand, training RBMs on non-decaying models should also, again, result in trained RBMs with decaying activation functions, except for the case of RBMs with Exponential hidden activation function. We will test this in numerical results reported below.\\
In numerical results reported in the following, training is performed by first defining a \textit{ground truth} model with distribution $p_{gt}(\boldsymbol{v})$. This is the training target (a lattice gas model or a ground-truth RBM). In Figure \ref{fig:NR_1}, two ground-truth lattice gas models are defined, a decaying interaction model 

with probability distribution 
\begin{equation}
p_{gt}(\boldsymbol{v}) = \frac{1}{Z}\exp \left[\sum_i I^{gt}_{i} v_i + \sum_{i<j}I^{gt}_{ij}v_i v_j+ \sum_{i<j<k}I^{gt}_{ijk}v_i v_j v_k \right]
\label{dec_gt}
\end{equation}
and a non decaying (three-body) interaction model described by Eq.\eqref{dec_gt} with $I_{i}^{gt}=I_{ij}^{gt}=0$ and $I_{ijk}^{gt}\neq0$.
The interaction values $I^{gt}_{k_1,...,k_s}$ for each model are reported in the caption of Fig. \ref{fig:NR_1}.\\
Subsequently, a randomly initialized RBM with probability distribution $p_{RBM}$ is trained to match the ground-truth by maximizing the negative cross-entropy $-H(p_{RBM},p_{gt})=\langle \ln p_{RBM}\rangle_{gt}$ (in analogy with likelihood maximization) by gradient ascent, where $\langle \cdot \rangle_{gt}$ is the expectation over $p_{gt}$. Training is performed by exact enumeration of the cross-entropy and its gradient at every training step, using the analytical expression of $p_{gt}$ directly, rather than sampling from it to generate the training dataset. Hyper-parameters are specified at every training simulation and reported in the figures caption. Unless otherwise stated, $c_{\mu}$ is set to 0, $\forall \mu$. The performance at the end of training is estimated by the ratio of the Kullback-Leibler $(KL)$ divergence between $p_{RBM}$ and $p_{gt}$ relative to the ground-truth entropy 
\begin{equation}
    \Delta_{KL}(p_{RBM}, p_{gt})= \frac{D_{KL}(p_{RBM}, p_{gt})}{H(p_{gt})},
\end{equation} 
where $D_{KL}(p_{RBM},p_{gt})=H(p_{RBM}, p_{gt})-H(p_{gt})$ and $H(p_{gt})= -\left\langle \ln p_{gt} \right\rangle_{gt}$.\\
Fig.\ref{fig:NR_2} shows the training process for the ground truth decaying model in Fig.\ref{fig:NR_1} and Eq.\eqref{dec_gt}.

\begin{figure}[htbp]
\centering
\includegraphics[width=\textwidth]{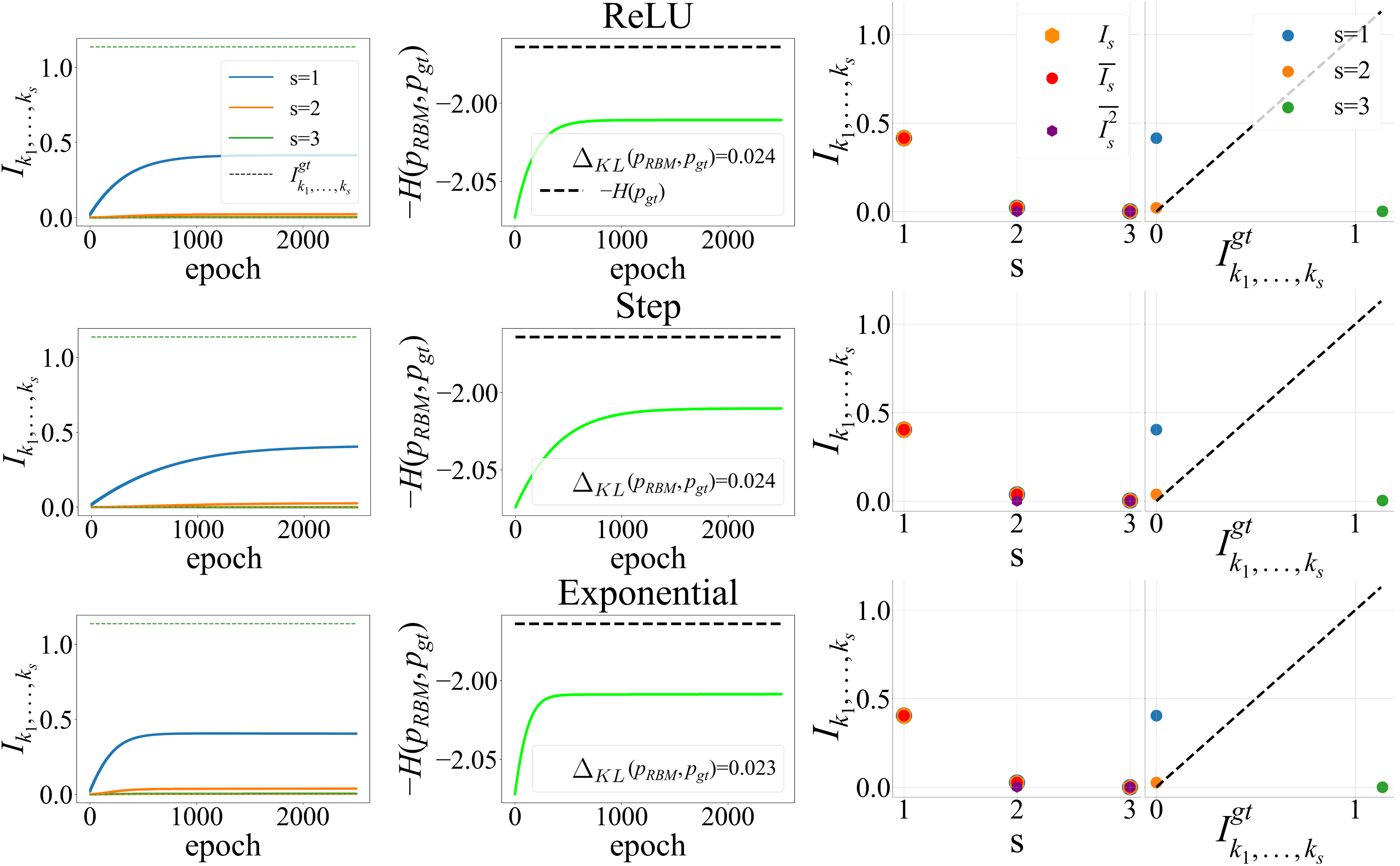}
\caption{\textbf{Learning a three body interaction model.} An RBM with $N=3$ and $M=4$, initialized with zero-mean Gaussian weights ($\sigma=0.01$), is trained to match the probability distribution of the non decaying model from Figure \ref{fig:NR_1} for different activation functions. The model is trained for 2500 epoch with a learning rate of 0.001. The first panel in each row shows the training trajectory of the interactions mapped from the RBM, compared with the ground truth interactions (dashed line). The second panel in each row shows the cross-entropy trajectory, where the target is the ground truth entropy (dashed line). {\red $\Delta_{KL}(p_{RBM},p_{gt})$ is reported in the legend for the RBM at the end of training.} The third panel in each row shows the interactions mapped from the trained RBM, with empirical and expected moments. The fourth panel in each row shows the comparison between ground truth and trained interactions.}
\label{fig:NR_3}
\end{figure}

In this case, the negative cross-entropy is properly maximized, as shown by its trajectory reaching the target for all the activation functions. The qualitative structure of the learned model closely resembles the ground-truth model. The trained interactions mapped from the RBM are relatively close to the ground truth, but, despite the very good cross-entropy value, they fail to reconstruct the ground-truth interactions exactly. In particular, the fields are overestimated compared to the ground-truth, leading to a compensation in the higher-order interactions. While for ReLU and Step the distance between the trained and ground truth interactions is quite large, for Exponential this distance stays smaller, in the tested case.\\
Figure \ref{fig:NR_3} shows the training process when the ground-truth model is the three-body interaction model in Figure \ref{fig:NR_1} and Eq.\eqref{dec_gt} with $I_{i}^{gt}=I_{ij}^{gt}=0$ and $I_{ijk}^{gt}\neq0$.

\begin{figure}[htbp]
\centering
\includegraphics[width=10cm]{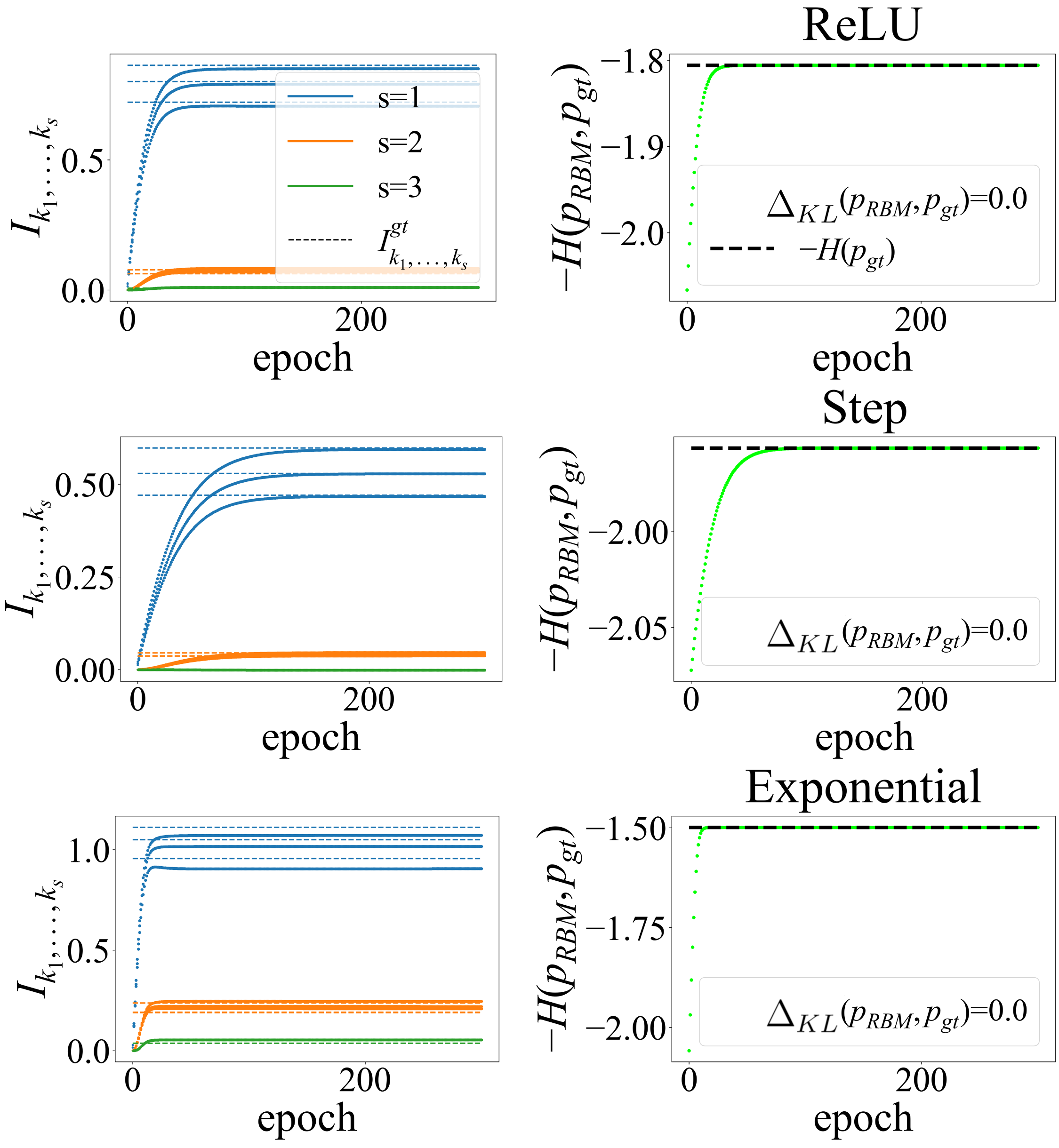}
\caption{\textbf{Learning interactions from a random RBM}. An RBM with $N=3$ and $M=4$, initialized with zero-mean Gaussian weights ($\sigma=0.01$), is trained to match the probability distribution of a Gaussian random RBM ($w_0=${\red $0.2$} and $g=${\red $0.2/\sqrt{M} $}) for different activation functions. The model is trained for 300 epoch with a learning rate of 0.02. The first panel for each activation function shows the training trajectory of the interactions mapped from the RBM, compared with the ground truth interactions (dashed line). The second panel for each activation function shows the cross-entropy trajectory, where the target (dashed line) is the ground truth entropy. {\red $\Delta_{KL}(p_{RBM},p_{gt})$ is reported in the legend for the RBM at the end of training.}}
\label{fig:NR_random}
\end{figure}

Here, the cross-entropy gets close enough to the entropy target ($\sim 2\%$ of the target), but farther away than in the decaying case. Clearly, the three body interaction is approximated by a combination of lower order interactions giving rise to a decaying interaction model. The absolute magnitude of pairwise and higher-order terms is extremely small compared to the one-body interactions. This means that the ground-truth model is learned as an independent model, where the fields set the average activation of each unit for all the activation functions.\\
Interestingly, both for the decaying and non-decaying models at the end of training, the empirical average is in very good agreement with the ensemble average computed in Eqs.\eqref{exp_exact} and \eqref{expected}, as shown in Figures \ref{fig:NR_2} and \ref{fig:NR_3}. This indicates that, rather than reconstructing the ground truth interactions, to which the algorithm has no direct access, the training trajectory leads to a regime in which the Gaussian approximation works effectively for the first moment and the cross-entropy value is very close to the ground truth entropy. If that is the case, then defining the ground truth as a random RBM with given $w_0$ and $g$, and learning its probability distribution, should yield a very good performance in terms of cross-entropy, and it should allow the exact reconstruction of the interactions from the mapped ground truth random RBM.\\ 
In fact, this is shown in Figure \ref{fig:NR_random}, where a ground-truth interaction model is built by applying Eq.\eqref{exact} to an RBM with Gaussian weights. An initialized RBM is then trained in the same way as in Figs.\ref{fig:NR_1} and \ref{fig:NR_2}.
The trained interactions are now very close to the ground-truth interactions for all the activation functions.

\begin{figure}[htbp]
\centering
\includegraphics[width=13.6cm]{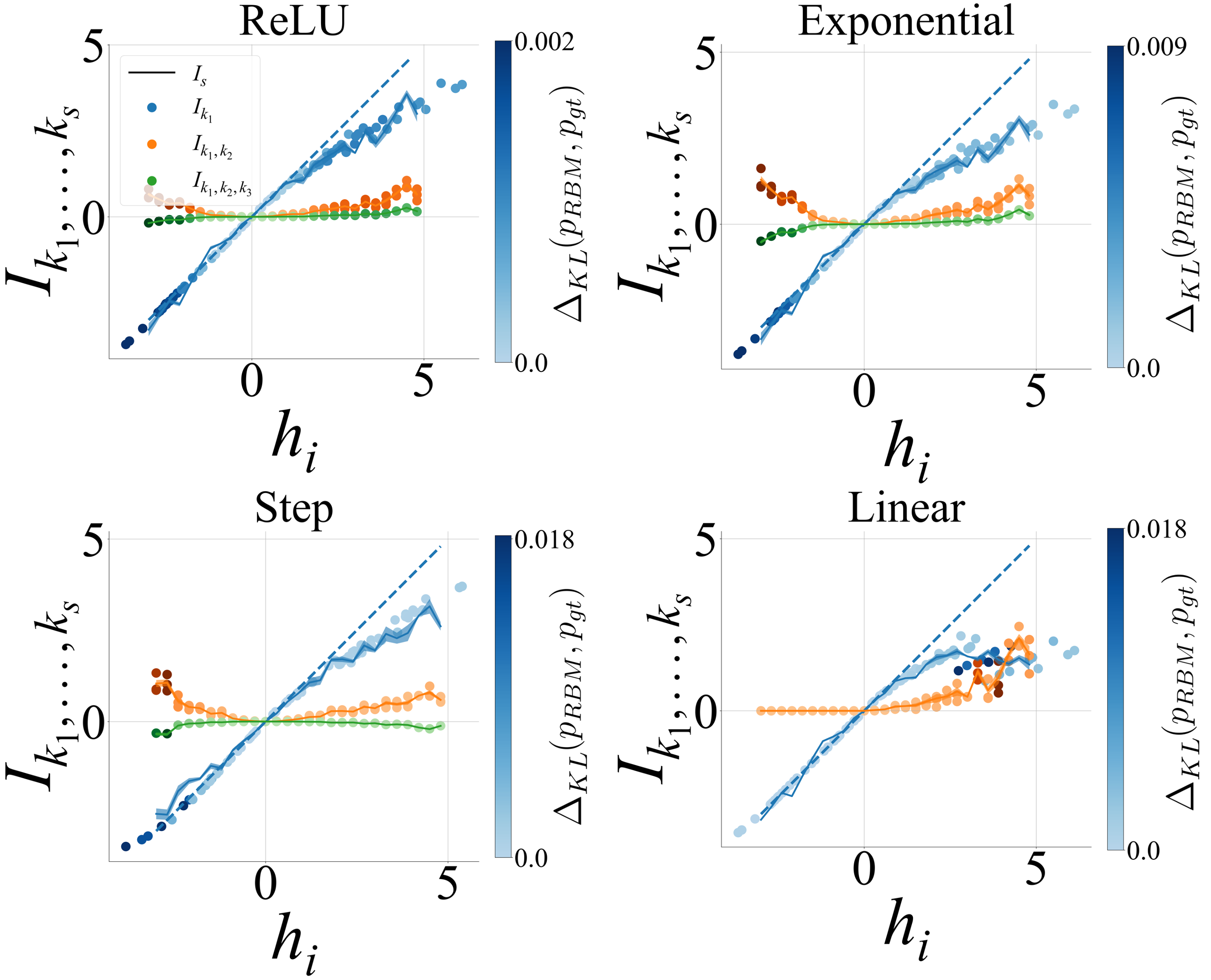}
\caption{\textbf{Learning an independent lattice gas model}. RBMs with $N = 3$ and $M =
4$, initialized with zero-mean Gaussian weights ($\sigma$ = 0.01), are trained to match the probability
distribution of ground truth models with one body interactions only (Eq.\eqref{field_only}) for different values of $h_i$. Interactions of order higher than 1 are plotted versus $h$. Each panel shows the comparison between the one-body ground truth interaction $h_i$ and the interaction terms $I_{k_1,...,k_s}$ of the trained model for a different activation function. Training is performed for 300 epochs with a learning rate of 0.02 (linear 0.3). {\red For each value of $h$, the colorbars show the $\Delta_{KL}$ value at the end of training.}}
\label{fig:NR_4}
\end{figure}

\begin{figure}[htbp]
\centering
\includegraphics[width=13.6cm]{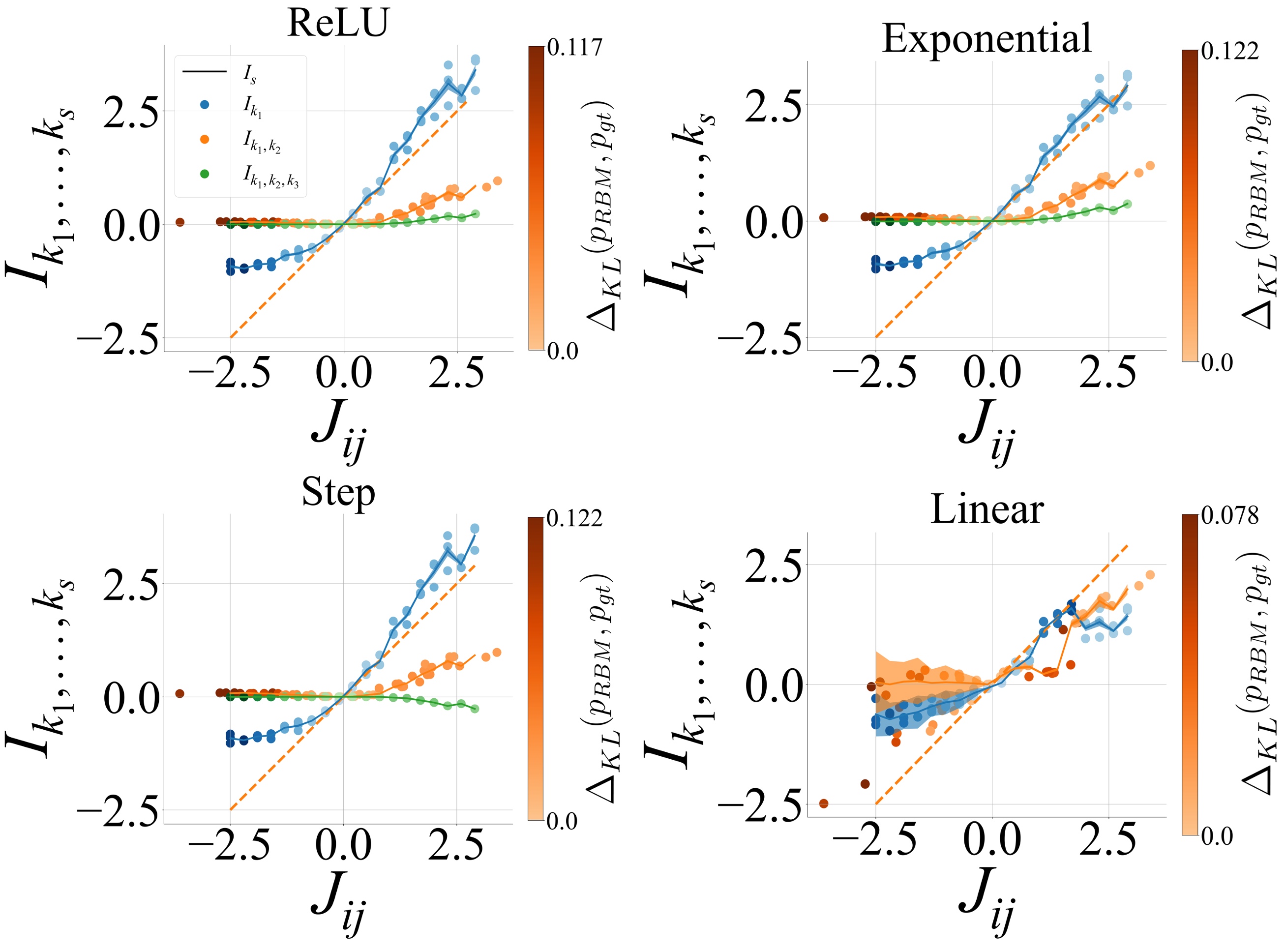}
\caption{\textbf{Learning a pairwise lattice gas model}. RBMs with $N = 3$ and $M =
4$, initialized with zero-mean Gaussian weights ($\sigma$ = 0.01), are trained to match the probability
distribution of ground-truth models with pairwise interactions only (Eq.\eqref{pair_only}) for different values of $J_{ij}$. Interactions of order 1 and 3 are plotted versus $J$. Each panel shows the comparison between the pairwise ground truth interaction $J_{ij}$ and the interaction terms $I_{k_1,...,k_s}$ of the trained model for a different activation function. Training is performed for 300 epochs with a learning rate of 0.02 (linear 0.3). {\red For each value of $J$, the colorbars show the $\Delta_{KL}$ value at the end of training.}}
\label{fig:NR_5}
\end{figure}

\begin{figure}[htbp]
\centering
\includegraphics[width=13.6cm]{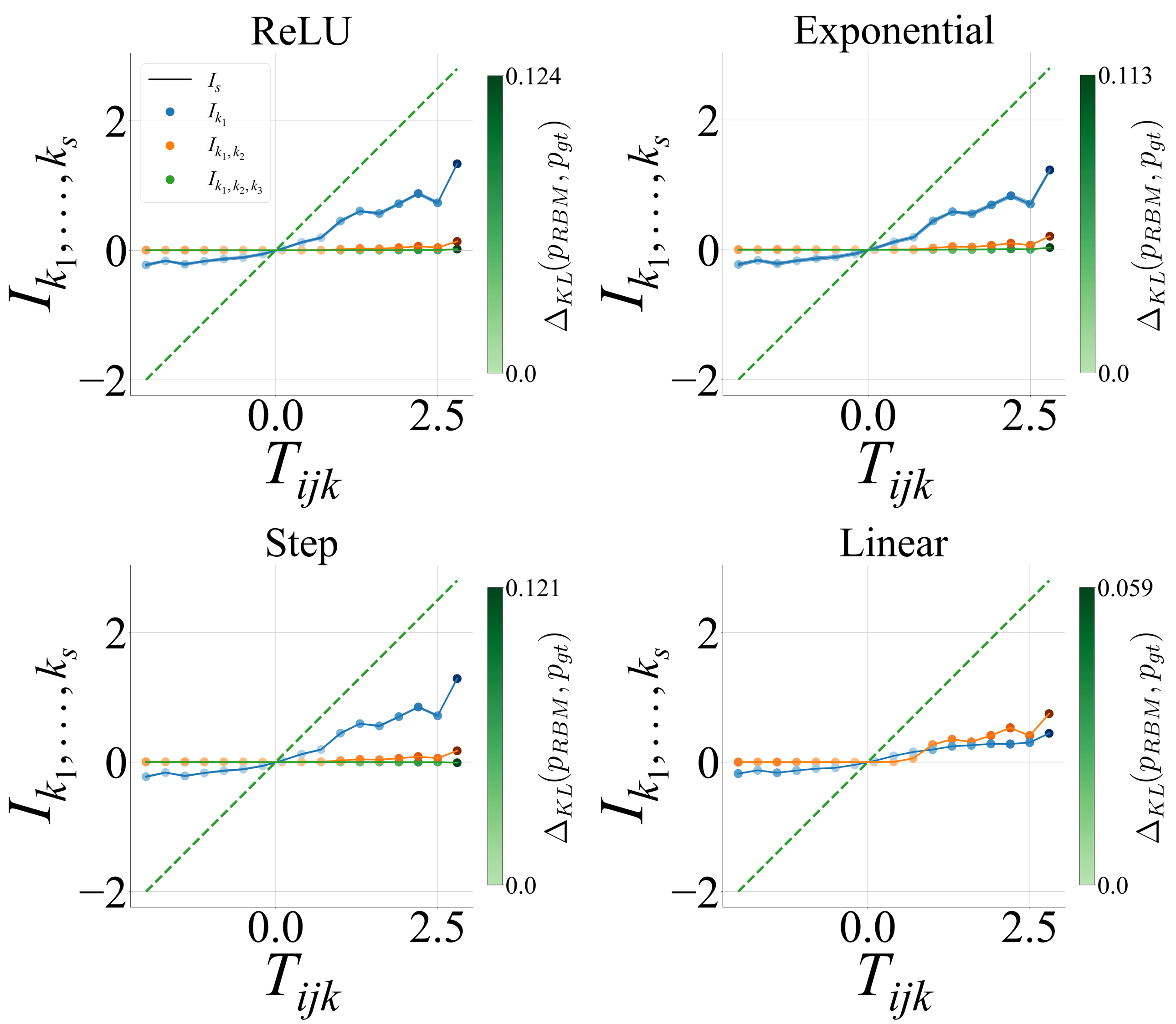}
\caption{\textbf{Learning a three-body lattice gas model}. RBMs with $N = 3$ and $M =
4$, initialized with zero-mean Gaussian weights ($\sigma$ = 0.01), are trained to match the probability distribution of ground truth models with one three-body interaction only (Eq.\eqref{three_only}) for different values of $T_{ijk}$. Interactions of $s<3$ are plotted versus $T$. Each panel shows the comparison between the three-body ground truth interaction $T_{ijk}$ and the interaction terms $I_{k_1,...,k_s}$ of the trained model for a different activation function. Training is performed for 300 epochs with a learning rate of 0.02 (linear 0.3). {\red For each value of $T$, the colorbars show the $\Delta_{KL}$ value at the end of training.}}
\label{fig:NR_6}
\end{figure}

To systematically inspect this behavior, we simulated the training process of an RBM learning an independent lattice gas model with varying field values for different activation functions.\\
The set of ground-truth lattice gas models to be learned is a set of one-body interaction models, where all higher-order interactions are zero 
\begin{equation}
p_{gt}(\boldsymbol{v}) = \frac{1}{Z}\exp\left[\sum_i h_i v_i \right],
\label{field_only}
\end{equation}
where $h_i\sim \mathcal{N}(h,h/5)$. These models are generated to avoid trivial sets of states, e.g. any model sampling only one state.\\
Parametrically sweeping the values of $h$ in the depicted range ($h\in [-3,5]$) ensures heterogeneity of each set of states. Figure \ref{fig:NR_4} shows how one-body interaction models in Eq.\eqref{field_only} are learned by an RBM, for different values of $h_i$ and different activation functions.
This kind of behavior suggests that the one-body interactions are correctly reconstructed by the trained RBM in the regime where $h$ is close to zero, with a satisfactory performance in terms of cross-entropy. When $h_i$ increases, the ground truth interactions are underestimated by the one-body interactions $I_{i}$, and this gets compensated by higher order terms, in order to minimize the cross-entropy. The empirical average is in very good agreement with the expected value on random RBM ensembles across different values of $h_i$.
The differences between activation functions in this case are very small, since the RBM weights stay relatively close to the initial conditions. Especially in the small $h$ regime, a small weight approximation (Eq.\eqref{pair_small}) shows that the behavior of different non-linearities gets similar to the linear case. Figure \ref{fig:NR_4_L} shows the same phenomenon for a larger network.\\
To check whether this behavior also happens for high-order interactions, we performed the same kind of analysis when the ground-truth models are pairwise and three-body
\begin{equation}
    p_{gt}(\bm{v}) = \frac{1}{Z}\exp\left[\sum_{i<j} J_{ij} v_i v_j \right] \label{pair_only}
\end{equation}
\begin{equation}
    p_{gt}(\bm{v}) = \frac{1}{Z}\exp\left[\sum_{i<j<k} T_{ijk} v_i v_j v_k \right] \label{three_only}
\end{equation}
where $J_{ij} \sim \mathcal{N} (J, J/5)$ and $T_{ijk} \sim \mathcal{N} (T, T/5)$. The results are shown in Figures \ref{fig:NR_5} and \ref{fig:NR_6}, respectively, and Figures \ref{fig:NR_5_L} and \ref{fig:NR_6_L} for larger networks.

These results show that regardless of whether the data come from a distribution with decaying or non-decaying interactions, the trained RBM exhibits a decaying interaction model that minimizes the cross-entropy, whose interaction structure is well described (in terms of averages) by a Gaussian ensemble of RBM weights. Note that, as shown in Appendix \ref{appD}, the three-body interaction that the RBM fails to reconstruct at the end of the learning process can indeed be potentially represented by the machine, in the sense that solutions (that is weights and biases) that lead to such interaction models do exist. However, such non-decaying solutions are not found through the learning process, likely because, in the weak coupling regime, they are very rare compared to the "sea" of decaying models. 

\begin{figure}[htbp]
\centering
\includegraphics[width=12cm]{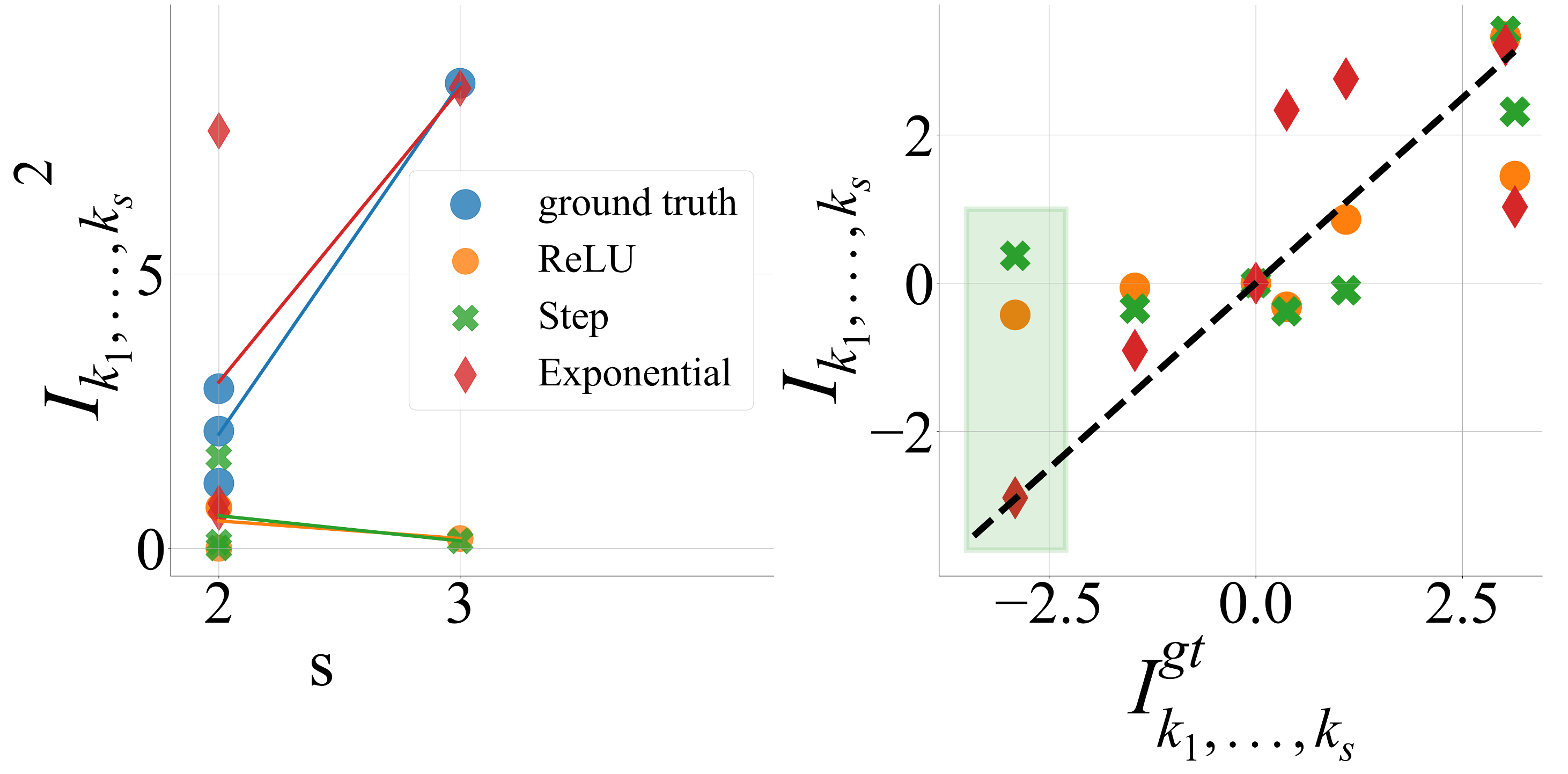}
\caption{{\red \textbf{Learning a non decaying lattice gas model with the Exponential activation}. An RBM with $N = 3$ and $M =
8$, initialized with Gaussian weights ($w_0=0.3$ and $g=3$), is trained on a ground truth non-decaying model. The left panel shows ${I^{gt}_{k_1,\cdots, k_s}}^2$ (in blue) and ${I_{k_1,\cdots, k_s}}^2$ from Eq.\eqref{exact} for the trained RBM with different activation functions. The lines connect Eq. \eqref{empirical} ($n=2$) for $s=2$ and $s=3$. The right panel shows the comparison of interaction terms for different activation functions between ground truth and trained RBM. The green patch highlights the three body interaction terms.}  }
\label{fig:NR_7}
\end{figure}

As noted in the previous section, in the strong coupling regime and with the Exponential activation function, one expects a different pattern, that is comparatively fewer decaying models in a "sea" of non-decaying one. We would thus expect that in this regime, the RBM with Exponential activation function can successfully learn a non-decyaing model, while the other activations functions fail to do so. We thus initialized an RBM close to the non-decaying regime for the Exponential activation ($w_0=0.3$ and $g=3$) and we trained it on a ground truth non-decaying model. Figure \ref{fig:NR_7} shows how the RBM with Exponential activation function is more successful in reconstructing the interaction terms of the ground truth, in particular, preserving the non-decaying nature of the model. The same does not hold for Step and ReLU that, also in this regime, learn the ground truth as a decaying interaction model. Moreover, in this case, the performance in terms of $\Delta_{KL}$ is much better for the Exponential activation function than for Step and ReLU; see Figure \ref{fig:NR_7_supp}.

\section{Discussion}
Our findings integrate into the major effort to develop a solid theory of the representational capacity and generative abilities of RBMs \cite{diSarra_2025,10.1007/978-3-642-33275-3_2,decelle2021restricted, e23010034,fachechi2025fundamental,decelle2024inferring} by describing the significant role of the form of the hidden-layer activation function. We first analyzed the statistics of interactions that emerge from random RBM ensembles with different activation functions. We demonstrated that the choice of activation function significantly influences the kinds of interaction model that the RBMs in such an ensemble are likely to represent. When Linear or Exponential activation functions are used, the mathematical structure of the equations becomes simpler, allowing the expected value and variance of interaction terms to be computed exactly over an ensemble of RBMs with Gaussian weights and biases. Yet, while in the Linear case the only non-zero interactions are the fields acting on visible units and the pairwise interactions between them, the Exponential case exhibits a markedly richer interaction structure, characterized by the emergence of higher-order terms. Specifically, in the latter case, we found that the expected value of the interaction terms can increase with the interaction order $s$. Moreover, the conditions under which the interaction fluctuations become larger than the interaction expected value show that higher order interactions are dominated by fluctuations in a larger parameter region, compared to lower order interactions.

For Step and ReLU, a small fluctuations expansion determines the interaction statistics in an approximate way. The same expansion gives exact results for the Linear RBM. Deterministic terms, expected values and variances for different interaction orders and activation functions are analyzed and compared with empirical statistics over a single RBM, showing a very good agreement between theoretical and empirical values. This agreement is broken for large $g$ when the values are approximated and holds in the entire parameter space when the values are computed exactly.

A key result of our study is that RBMs with rapidly increasing activation functions, such as the Exponential function, present a regime in which many models in the ensemble exhibit higher order interactions with larger magnitude than low order ones, and the interaction models are thus mostly non-decaying. This behavior suggests that when RBMs are trained on data, such activation functions may enable the model to better capture high-order correlations. The parameter regions in which this feature occurs are analytically determined to be a large $w_0$ region and a region where $w_0$ is small but the variance of the weight distribution in the ensemble is large. The latter is easily accessible during training and facilitates the representation of high-order interactions with the Exponential activation function. Conversely, activation functions like ReLU and Step lead to models where lower-order interactions are likely to remain dominant.

This result may be relevant in light of recent generalizations of the Hopfield model exhibiting substantially improved storage and retrieval properties \cite{AGLIARI2026108181,Agliari2025HeteroAssociative}. When viewed as interacting systems, these architectures effectively introduce correlations that extend beyond the pairwise interactions characterizing the standard Hopfield model. Since the latter is known to correspond to an RBM with linear hidden units, one may expect that representing these generalized associative-memory models within the RBM framework would require hidden units whose activation functions induce higher-order interactions among visible units. Our finding that learning non-decaying models is generally difficult, except in some limited cases with Exponential activation function, would also mean that learning such models from data may be difficult unless regularizations are used \cite{Ventura_2024}. This remains to be addressed in future studies.

Our theoretical analysis relies on the choice of a random RBM ensemble, where each RBM has Gaussian \textit{i.i.d.} weights, and exact computations can be easily performed. We used this to gain insights into the nature of RBMs prevalent in a given ensemble (decaying versus non-decaying). These insights allowed to draw conclusions about what types of models are more likely to emerge from the process and tested this in a number of simple examples. We are, however, mindful that learning is a highly complex and structured process. Consequently, analyzing the learning dynamics and different learning regimes, as has been done in \cite{fachechi2025fundamental} for the Linear activation function, for other activation functions would be a crucial step to make conclusive statement.

Previous work suggested that the great generalization ability in neural networks lies in the emergence of a 'simplicity bias' during learning. This is the tendency of the machine to first learn low-order features and only later fit high-order, more complex statistics \cite{shah2020pitfalls,rende2024distributional} and it has been argued to emerge from stochastic gradient descent \cite{refinetti2023neural}. Our results suggest that a 'simplicity bias' can also arise from the representational bias of RBMs introduced by hidden units activation function, where low-order interactions are likely to be dominant compared to high-order ones. 

A natural step from here would be to extend the present analysis to alternative statistical ensembles. Such extensions would not only test the robustness of our conclusions beyond Gaussian ensembles, but also uncover representational properties that are not directly accessible under the Gaussian assumption. More broadly, these results emphasize the importance of activation functions as a design choice in neural network architectures. By tailoring activation nonlinearities to the statistical structure of the data, it may be possible to promote the extraction of the most informative representations for a given task. Similarly, it would be interesting to see how the choice of activation functions influences other aspects of the functioning of RBMs. And example of such problems is how the choice of activation functions can affect the outcome of closed-loop learning, the resulting decline in performance and model collapse under retraining. Model collapse is a problem of much current interest \cite{shumailov2024ai} and the RBM appears to provide a platform for testing theoretical findings in simple models to a more complex setting \cite{Jangjoo}. Describing which distributions are more likely to be represented by a given architecture, as we did for RBM in this paper, can shed light on the reduction or increase of data diversity occurring through retraining, and possible ways of preventing it.

\section*{Acknowledgments}
We are grateful for interesting discussions, support and contribution throughout the development of the present work to Nicola Bulso. The study was supported by Research Council of Norway Centre for Neural Computation, grant number 223262 (GdS and YR); Research Council of Norway Centre for Algorithms in the Cortex, grant number 332640 (GdS and YR); Research Council of Norway Centre NORBRAIN, grant number 295721 (GdS and YR); The Kavli Foundation (GdS and YR). The funder had no role in study design, data collection and analysis, decision to publish, or preparation of the manuscript.

\section*{Data availability}
Data and code are available at \url{https://github.com/gdisarra/RBM_nonlinearity}.

\section*{Appendix}
\appendix

\subsection{The expected interaction in the Linear case}
\label{appA}
Defining $n\equiv s-p$
\begin{eqnarray}
\left \langle I_{k_1,\cdots, k_s}\right \rangle &=& \sum_{\mu} \sum^{s}_{n=1} (-1)^{s-n} \sum_{1\leq j_1<j_2\cdots < j_n\leq n} \frac{1}{2}\left\{\sum^{n}_{l=1} \langle (w_{k_{j_l},\mu})^2\rangle +\sum^{n}_{l\neq l'=1}\langle w_{k_{j_l},\mu} \rangle \langle w_{k_{j_{l'}},\mu} \rangle\right\}\\
&-& \sum_{\mu} c_{\mu}\sum^{s}_{n=1} (-1)^p \sum_{1\leq j_1<j_2\cdots < j_n\leq n} \left\{\sum^{n}_{l=1} \langle w_{k_{j_l},\mu}\rangle\right\}
\label{eq:lin_exact1}
\end{eqnarray}
Defining $w_0\equiv \langle w_{k_{j_l},\mu} \rangle$ and $w_2\equiv \langle w_{k_{j_l},\mu}^2\rangle$, we can write this as 
\begin{eqnarray}
\left \langle I_{k_1,\cdots, k_s}\right \rangle &=& \sum_{\mu} \sum^{s}_{n=1} (-1)^p \sum^{s}_{1\leq j_1<j_2\cdots < j_n\leq n} \frac{1}{2}\left\{\sum^{n}_{l=1} w_2 +\sum^{n}_{l\neq l'=1} w^2_0\right\}\\
&-& \sum_{\mu} c_{\mu}\sum^{s}_{n=1} (-1)^{s-n} \sum^s_{1\leq j_1<j_2\cdots < j_n\leq n} \left\{\sum^{n}_{l=1} w_0\right\}\\
&=&  \sum_{\mu} \sum^{s}_{n=1} (-1)^{s-n} \binom{s}{n} (nw_2/2+n(n-1)w^2_0/2) - \sum_{\mu} c_{\mu}\sum^{s-1}_{p=0} (-1)^p \binom{s}{n} (n w_0)\\
&=& \sum_{\mu} \sum^{s}_{n=1} (-1)^{s-n} \binom{s}{n} K_{\mu}(nw_0)+ \sum_{\mu} \sum^{s}_{n=1} (-1)^{s-n} \binom{s}{n} n (w_2-w^2_0)/2
\label{eq:lin_exact_deriv}.
\end{eqnarray}
Noting that $\sum^{s}_{n=1} (-1)^{s-n} \binom{s}{n} n=0$ for $s\ge2$ and that $\sum^{s}_{n=1} (-1)^{s-n} \binom{s}{n} n^2=\delta_{s,2}$ we find the expression in Eq.~\eqref{eq:lin_exact}.

\subsection{Derivation of Eq.~\eqref{exp_exact2}}
\label{app_expcorr}
\begin{eqnarray}
\langle I^{\rm Exp}_{k_1,\cdots,k_2}I^{\rm Exp}_{k'_1,\cdots,k'_2}\rangle 
&=& \sum_{\mu,\nu} e^{-c_{\mu}-c_{\nu}} \prod^{s}_{i=1}\prod^{s'}_{j=1}  \langle (\exp w_{k_i\mu}-1) (\exp w_{k'_j\nu}-1)\rangle \\
&=& \sum_{\mu} e^{-2c_{\mu}}  \langle (\exp w_{i\mu}-1)^2 \rangle^m \gamma^{s+s'-2m}_1+\sum_{\mu\neq \nu} e^{-c_{\mu}-c_{\nu}} \prod^{s}_{i=1}\prod^{s'}_{j=1}  \langle (\exp w_{k_i\mu}-1)\rangle \langle(\exp w_{k'_j\nu}-1)\rangle\\
&=& \sum_{\mu} e^{-2c_{\mu}} \gamma_2^m \gamma^{s+s'-2m}_1+\sum_{\mu\neq \nu} e^{-c_{\mu}-c_{\nu}} \gamma^{s+s'}_1 = \gamma^{s+s'}[\gamma^m_2\gamma^{-2m}_1\sum_{\mu} e^{-2c_{\mu}}+\sum_{\mu\neq \nu} e^{-c_{\mu}-c_{\nu}}]
\end{eqnarray}
Subtracting $\langle I^{\rm Exp}_{k_1,\cdots,k_2}\rangle \langle I^{\rm Exp}_{k'_1,\cdots,k'_{s'}}\rangle = \gamma^{s+s'}_1 \sum_{\mu,\nu} \exp^{-c_{\mu}-c_{\nu}}$ yields Eq.~\eqref{exp_exact2}.

\subsection{Expansion in Eq.~\eqref{fluct_comp}}
\label{appB}

We can prove expression Eq.~\eqref{fluct_comp} by noting that 
\begin{equation}
    \sum_{1\leq j_{1}<j_{2}<...<j_{n}\leq n}^{s} \sum^{s}_{l=1} x_{k_{j_{l}}}=\binom{s-1}{n-1} \sum^{s}_{l=1} x_{k_l}
    \label{eq:proof}
\end{equation}
To see this, consider the the case of the number of times $x_{k_{s}}$ appears in the above sum. This is the terms where $j_n=s$-- as for $l<n$, $j_l$ will have a maximum $s-n+l$--  and will be the number of terms in  $\sum_{1\leq j_{1}<j_{2}<...<j_{n-1}\leq n-1}^{s-1}$, which is $\binom{s-1}{n-1}$. $x_{k_{s-1}}$ appears when $j_n=s-1$, that is $\binom{s-2}{n-1}$ times (using the same logic as the the first case but $s$ replaced with $s-1$), or $j_{n-1}=s-1$, or when $j_n=s$ that is $\binom{s-2}{n-2}$ (using the same logic as the first case but with $s$ and $n$ replaced with $s-1$ and $n-1$), summing to $\binom{s-1}{n-1}$, and so on for any $x_{k}$. Using $\delta w_{k_{j_l}}$ in place of $x_{k_l}$ in Eq.~\eqref{eq:proof} yields the expression in Eqs.~\eqref{alpha} and using $\delta w_{k_{j_l}}^2$ in Eq.~\eqref{eq:proof}, yields Eqs.~\eqref{beta} for $\alpha$ and $\beta$ in Eq.~\eqref{fluct_comp}. 

The expression for $\eta$ in Eq.~\eqref{eta} can be similarly derived by showing that 
\begin{equation}
    \sum_{1\leq j_{1}<j_{2}<...<j_{n}\leq n}^{s} \sum_{m\neq l=1}^{n}x_{k_{j_{l}}} x_{k_{j_{m}}} = \binom{s-2}{n-2} \sum_{m<l} x_{k_{j_{l}}} x_{k_{j_{m}}}
    \label{eq:proof2}
\end{equation}
To see this, first consider the case of the number of times $x_{k_{s}}x_{k_{s-1}}$ appears in this sum, which is the number of times $j_n=s$ and $j_{n-1}=s-1$. This would be the number of terms in  $\sum_{1\leq j_{1}<j_{2}<...<j_{n-2}\leq n}^{s-2}$ namely $\binom{s-2}{n-2}$ as shown above. The number of times $x_{k_{s}}x_{k_{s-2}}$ appears in the sum can be shown to be the same, as it is equal to the number of times $j_n=s$, $j_{n-1}=s-2$, that is $\binom{s-3}{n-2}$ (as per the first case with $s$ replaced with $s-1$), plus the number of times $j_n=s$, $j_{n-1}=s-1$ and $j_{n-2}=n-2$, that is $\binom{s-3}{n-3}$ (as per the first case with $s$ and $n$ replaced with $s-1$ and $n-1$). These again sum to $\binom{s-2}{n-2}$. 

\subsection{Computation of the covariance}
\label{appC}

The covariance can be computed considering two sets of different indexes $\boldsymbol{k}=\{ k_{i} \}_{i=1,..,s}$ and $\boldsymbol{k}'=\{ k'_{i} \}_{i=1,..,s}$.\\
Given two random variables $\delta I_{k_{1},...,k_{s}}$ and $\delta I_{k'_{1},...,k'_{s}}$, generated by the same probability distribution $P(\delta I_{k_{1},...,k_{s}}) = P(\delta I_{k'_{1},...,k'_{s}})$, we want to quantify their joint variability, that is their covariance. \\
This will obviously depend on the index choice: the number of equal indexes ($k_{i}=k'_{j}, \; \forall i,j$) in the two sets will change the covariance value.\\
Let us first address the problem
\begin{equation}
    \left\langle \delta I_{k_{1},...,k_{s}} \delta I_{k'_{1},...,k'_{s}} \right\rangle \qquad \boldsymbol{k} \cap  \boldsymbol{k}'= \emptyset
    \label{disjoint}
\end{equation}

where the interaction terms are involving disjoint sets of visible nodes. Then
\begin{eqnarray*}
    \left\langle \delta I_{k_{1},...,k_{s}} \delta I_{k'_{1},...,k'_{s}} \right\rangle
    & = & \sum_{\mu,\nu}\beta_{s\mu}\beta_{s\nu}\sum_{i,j}^{s} \left\langle \delta w_{k_{i}\mu}^{2}\right\rangle \left\langle \delta w_{k'_{j}\nu}^{2}\right\rangle\\
    & = & \left[\frac{sg^{2}}{M} \sum_{\mu}\beta_{s\mu}\right]^{2}\\
    & = & \left\langle \delta I_{k_{1},...,k_{s}}\right\rangle^{2}
\end{eqnarray*}
We can conclude, as expected, that two interaction terms sharing no visible nodes are uncorrelated and independent.\\
A different scenario is shown when interaction terms are not involving disjoint sets of visible nodes, meaning that there exist at least one element $k_{i} \in \boldsymbol{k}$ such that $k_{i}=k'_{j}$, $\forall i,j$ with $k'_{j}\in \boldsymbol{k}'$.\\
We can define
\begin{equation*}
    Q \equiv \boldsymbol{k} \cap \boldsymbol{k}' \neq \emptyset \qquad Q=\{k_{i}\}_{i=1,...,q}
\end{equation*}
where $Q$ is the set of visible nodes involved in both interaction terms and $q$ is the cardinality of $Q$. 
In this case, a crucial role is played by $q$. This happens because sharing a visible node means summing over the same columns in the weight matrix. We can show this with a simple example.\\
Let's consider two interaction terms sharing the visible node $k_{1}$ and no other, $k_{i}\neq k'_{j}\; \forall i\neq1,j$:
\begin{equation}
    \delta I_{k_{1},k_{2},...,k_{s}} \propto \sum_{i}^{s}\sum_{\mu}\delta w_{k_{i}\mu} = \sum_{\mu} \delta w_{k_{1}\mu} + \sum_{i\neq 1}^{s}\sum_{\mu} \delta w_{k_{i}\mu} 
    \label{ex1}
\end{equation}
\begin{equation}
    \delta I_{k_{1},k'_{2},...,k'_{s}} \propto \sum_{i}^{s}\sum_{\mu}\delta w_{k'_{i}\mu} = \sum_{\mu} \delta w_{k_{1}\mu} + \sum_{i\neq 1}^{s}\sum_{\mu} \delta w_{k'_{i}\mu}
    \label{ex2}
\end{equation}
where the first sum on the r.h.s. of the relation runs exactly over the same weights in the two interaction terms. The covariance tells us how much this kind of terms impacts in the relationship between two interaction terms, involving different sets of visible nodes through the value $q$. If $q=0$, condition~\eqref{disjoint} is satisfied, and between equations~\eqref{ex1} and~\eqref{ex2} there are no shared terms, so that the sums run over different elements in the weight matrix. It follows immediately from gaussian independence, that the interaction terms are independent, as proved earlier. If $q=s$, the same set of visible nodes is involved in both interaction terms, the covariance is maximal and corresponds to the variance $\left\langle {I_{k_{1},...,k_{s}}}^{2}\right\rangle-\left\langle I_{k_{1},...,k_{s}}\right\rangle^{2}$.\\
Then we can compute $\left\langle \delta I_{k_{1},...,k_{s}} \delta I_{k'_{1},...,k'_{s}} \right\rangle$ for $q\in [1,s-1]$:
\begin{eqnarray*}
     \left\langle \delta I_{k_{1},...,k_{s}} \delta I_{k'_{1},...,k'_{s}} \right\rangle &
    = & \frac{g^{2}}{M}\sum_{\mu}\left[ q{\alpha_{s\mu}}^{2} + \frac{g^{2}}{M}\left( 2q{\beta_{s\mu}}^{2} + s^{2} \beta_{s\mu}\sum_{\nu} \beta_{s\nu} + {\gamma_{s\mu}}^{2}\frac{q(q-1)}{2}\right)\right]\\
    &=& \frac{g^{2}}{M}\sum_{\mu} q{\alpha_{s\mu}}^{2} + \mathcal{O}(\delta w^3)
\end{eqnarray*}
From this expression we can retrieve both the independent case, when two disjoint sets are considered ($q=0$)
\begin{equation*}
    \left\langle \delta I_{k_{1},...,k_{s}} \delta I_{k'_{1},...,k'_{s}} \right\rangle\biggl|_{q=0} = \frac{(sg^{2})^{2}}{M^{2}} \sum_{\mu,\nu}\beta_{s\mu} \beta_{s\nu} =
    \left\langle \delta I_{k_{1},...,k_{s}}\right\rangle^{2}
\end{equation*}
and the variance, when the two sets are identical ($q=s$) 
\begin{eqnarray*}
    \left\langle \delta I_{k_{1},...,k_{s}} \delta I_{k'_{1},...,k'_{s}} \right\rangle\biggl|_{q=s} & = & \frac{sg^{2}}{M} \sum_{\mu} \left[ {\alpha_{s\mu}}^{2} + \frac{g^{2}}{M} \left(2 {\beta_{s\mu}}^{2} + s \beta_{s\mu} \sum_{\nu} \beta_{s\nu} + \frac{\left(s-1\right)}{2} {\gamma_{s\mu}}^{2}\right)\right] = \left\langle {\delta I_{k_{1},...,k_{s}}}^{2} \right\rangle    
\end{eqnarray*}
Then the final expression for the covariance is 
\begin{eqnarray*}
    cov(\delta I_{k_{1},...,k_{s}}, \delta I_{k'_{1},...,k'_{s}}) & = & \left\langle \delta I_{k_{1},...,k_{s}} \delta I_{k'_{1},...,k'_{s}} \right\rangle- \left\langle \delta I_{k_{1},...,k_{s}}\right\rangle^{2}\\
    \
    & = &\frac{q\sigma^{2}}{M^{2}}\sum_{\mu}\left[ {\alpha_{s\mu}}^{2} + 2{\beta_{s\mu}}^{2}\sigma^{2} + {\gamma_{s\mu}}^{2}\frac{q-1}{2}\sigma^{2}\right]
\end{eqnarray*}

\subsection{RBMs represent arbitrary three-body interaction model.}
\label{appD}
The statistical analysis in section \ref{sec3} describes the kind of interaction structures that stem from an ensemble of random Gaussian RBMs. These structures are largely characterized by decaying interaction models, except for a region in the parameter space of the RBM with Exponential activation. Deviations from the decaying structure can generally occur either from rare random fluctuations in single realizations of the Gaussian ensemble, or by the presence of correlations and dependencies in the RBM weights, which break the Gaussian \textit{i.i.d.} assumption. In fact, the full repertoire of interaction models that RBMs are able to represent is much larger than the decaying class. In principle, the training process is able to introduce correlations and dependencies that break the \textit{i.i.d.} Gaussian assumptions and allow for the representation of more complicated interaction structures. However, this does not always happen in practice.\\ In section \ref{sec4} (Figures \ref{fig:NR_3} and \ref{fig:NR_6}) we show that the RBM learns a particular kind of non-decaying interaction model, namely a pure three-body interaction, as a decaying interaction model that approximates it. 
This raises the question of understanding whether the kind of ground truth model that the RBM is trying to learn can be represented in the first place.\\ 
We check this for the three-body interaction model by numerically solving the non-linear set of equations given by Eq.\eqref{exact} for a subset of the weights of a small RBM weights. 
\begin{figure}[htbp]
\centering
\includegraphics[width=12cm]{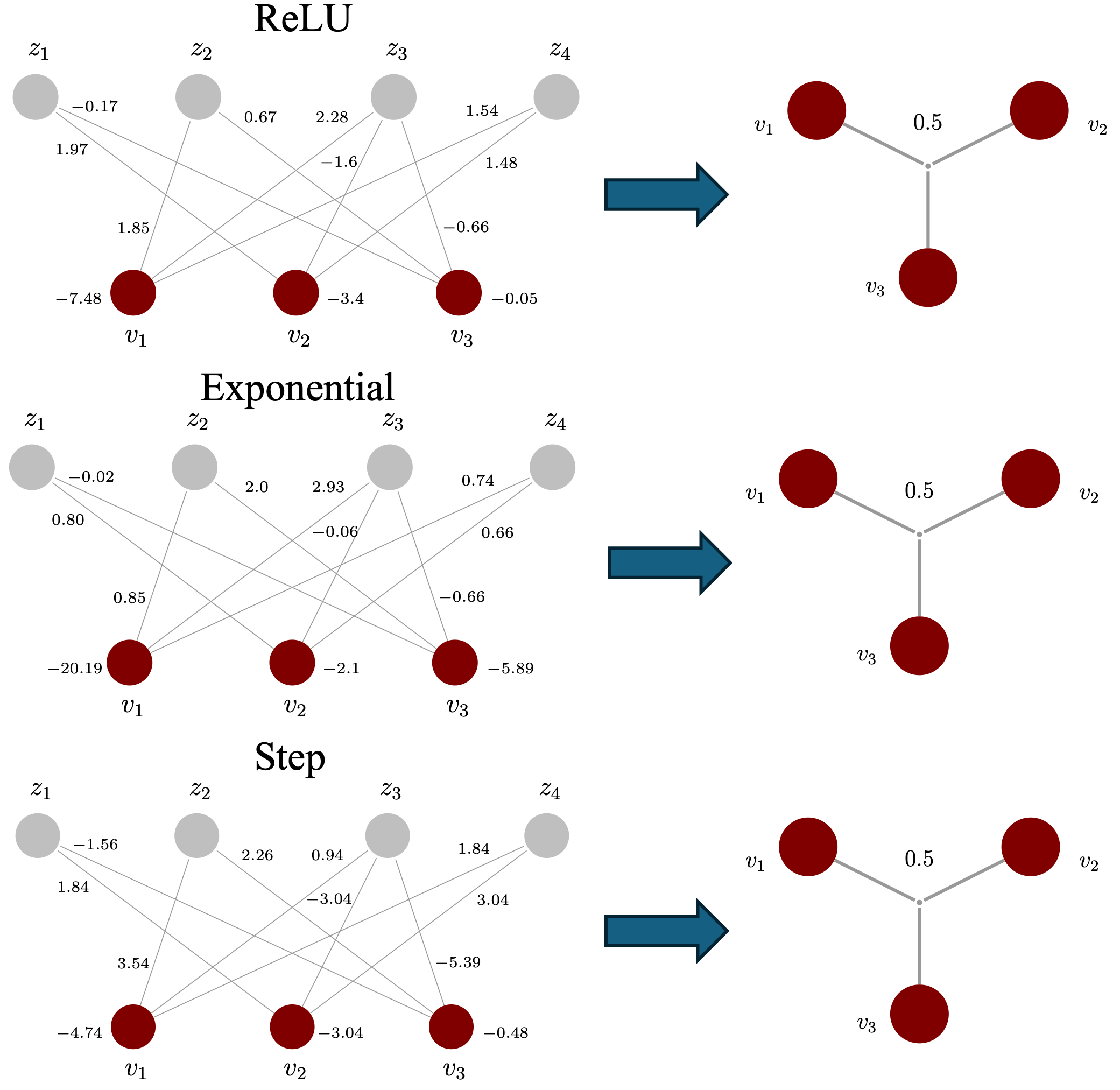}
\caption{RBMs represent three-body interaction model with $T=0.5$. The weights configurations solving the non-linear set of equations given by Eq.\eqref{exact} for a three-body interaction is shown for each activation function.}
\label{fig:three_rep}
\end{figure}

In particular, we generate an RBM with $N=3$ and $M=4$ and random weights and we set the pairwise interactions to zero by solving $I_{ij}=0$ and $I_{ijk}=T$ for a selected subset (four out of 12, 3 for the pairwise interactions and one for the three-body interaction) of the weights (Figure \ref{fig:three_rep} for $T=0.5$):
\begin{equation}
    \sum_{\mu}^M \left[ K\left( w_{i\mu}+w_{j\mu}\right)-K\left( w_{i\mu}\right)-K\left( w_{j\mu}\right)\right]=0.
\end{equation}
and 
\begin{equation}
    \sum_{\mu}^M \left[ K\left( \sum_{j}w_{k_j\mu}\right)-\sum_{j_1<j_2}K\left( w_{k_{j_1}\mu}+w_{k_{j_1}\mu}\right)+\sum_jK\left( w_{k_j\mu}\right)\right]=T.
\end{equation}
On the same RBM, we set the biases to satisfy $I_{i}(b_i) = 0$
\begin{equation}
     b_{i} =- \sum_{\mu}^M  K_{\mu}\left( w_{i\mu}\right)
\end{equation}
In this way, regardless of the activation function, we found a parameter configuration that generates a pure three-body interaction model with arbitrary interaction $T$.

\clearpage
\section*{Supplementary figures}
\begin{figure}[htbp]
\centering
\includegraphics[width=10cm]{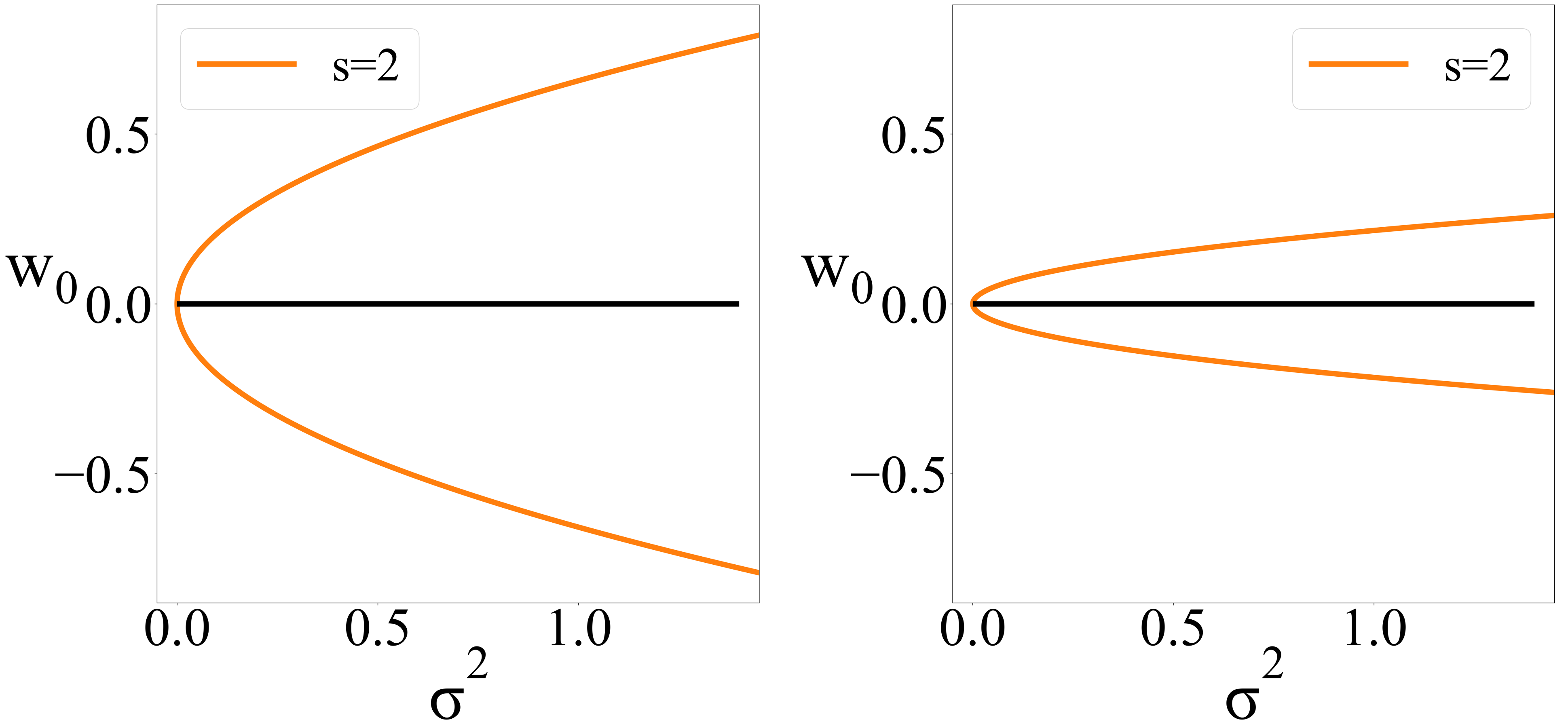}
\caption{Solutions of $\Delta^{\rm Lin}_{s}=1$ in the $(\sigma^2,w_0)$ plane for $M_0=0.1$ (left) and $M_0=0.002$ (right). Eq.~\eqref{delta_line} is plotted with a color corresponding to the order of interaction. The black line shows the divergence $w_0=0$, where interaction fluctuations are infinitely larger than their mean.}
\label{fig:lin_delta1}
\end{figure}
\begin{figure}[htbp]
\centering
\includegraphics[width=\textwidth]{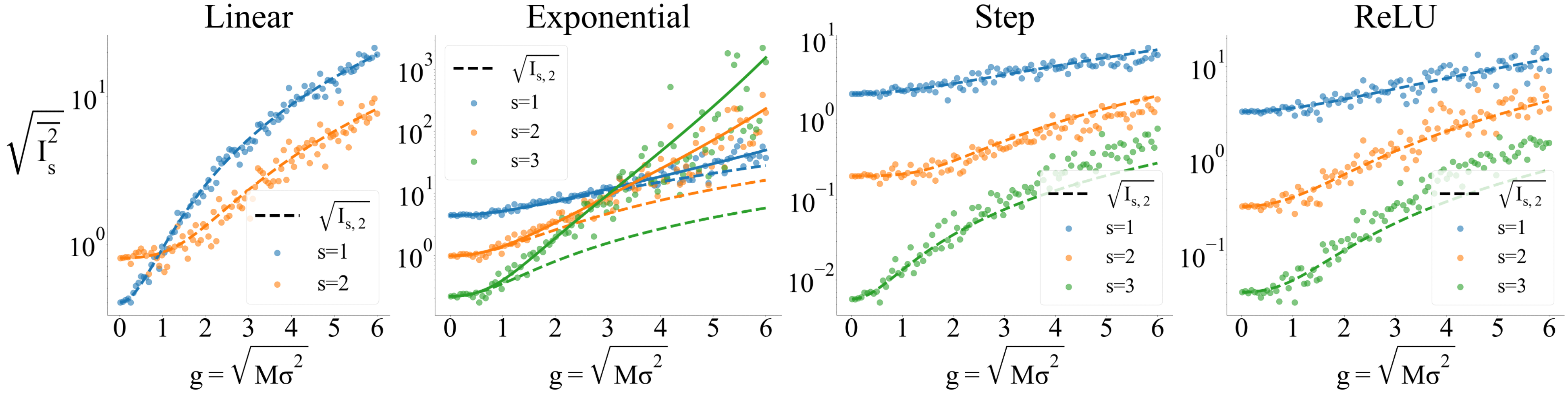}
\caption{Square root of $\overline{I_s^2}$ from Eq.~\eqref{empirical} ($n=2$) and square root of $I_{s,2}$ from Eq.~\eqref{moment2} (dashed line) versus $g$ for $w_0=0.2$. The solid line for the Exponential activation shows the first term in Eq.~\eqref{exp_exact2}. The RBM parameters are $b_i=0$ $\forall i$, $c_{\mu}=0$ $\forall \mu$, $N=8$ and $M=20$.}
\label{fig:I_mom2_g}
\end{figure}
\begin{figure}[htbp]
\centering
\includegraphics[width=\textwidth]{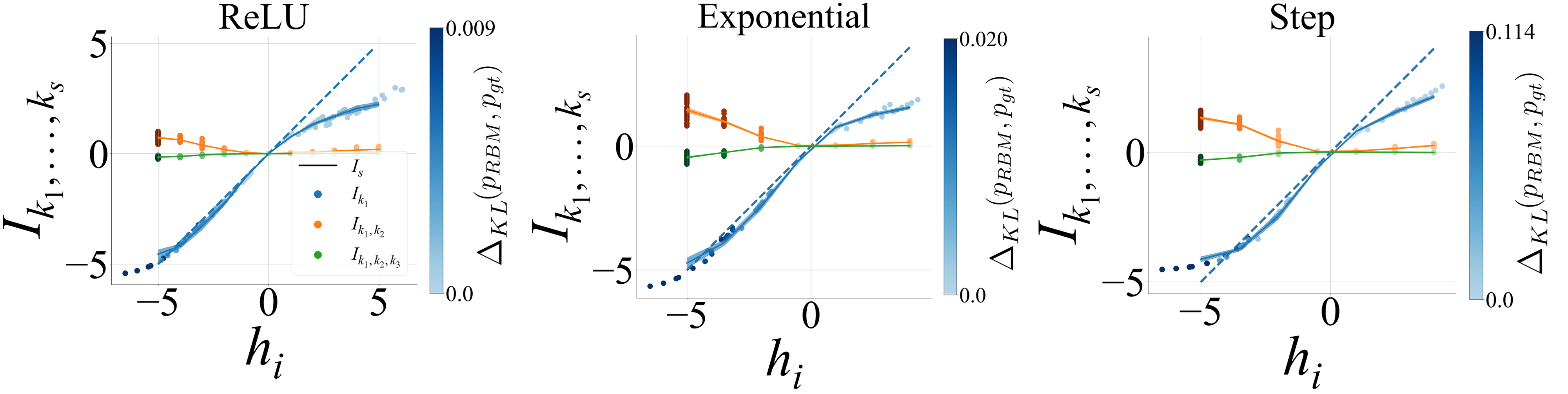}
\caption{Learning a large independent lattice gas model. RBMs with $N = 10$ and $M =
15$, initialized with zero-mean Gaussian weights ($\sigma$ = 0.01), are trained to match the probability
distribution of ground truth lattice gas models with one body interactions only (Eq.\eqref{field_only}) for different values of $h_i$ and for different activation functions. Each panel shows the comparison between the one-body ground truth interaction $h_i$ and the interaction terms $I_{k_1,...,k_s}$ of the trained model. The model is trained for 300 epoch with a learning rate of 0.008 (linear 0.3).}
\label{fig:NR_4_L}
\end{figure}

\begin{figure}[htbp]
\centering
\includegraphics[width=\textwidth]{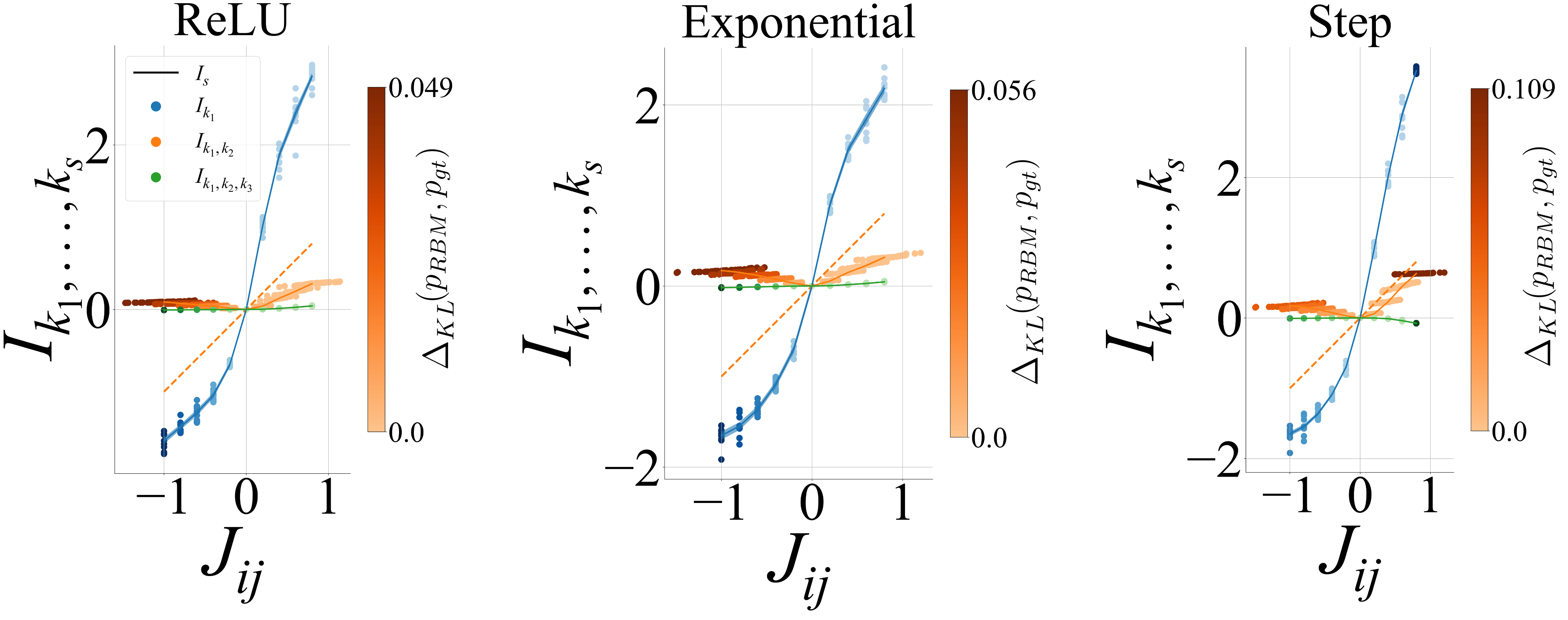}
\caption{Learning a large pairwise lattice gas model. RBMs with $N = 10$ and $M =
15$, initialized with zero-mean Gaussian weights ($\sigma$ = 0.01), are trained to match the probability
distribution of ground truth lattice gas models with pairwise interactions only (Eq.\eqref{pair_only}) for different values of $J_{ij}$ and for different activation functions. Each panel shows the comparison between the pairwise ground truth interaction $J_{ij}$ and the interaction terms $I_{k_1,...,k_s}$ of the trained model. The model is trained for 300 epoch with a learning rate of 0.008 (linear 0.3). }
\label{fig:NR_5_L}
\end{figure}

\begin{figure}[htbp]
\centering
\includegraphics[width=15cm]{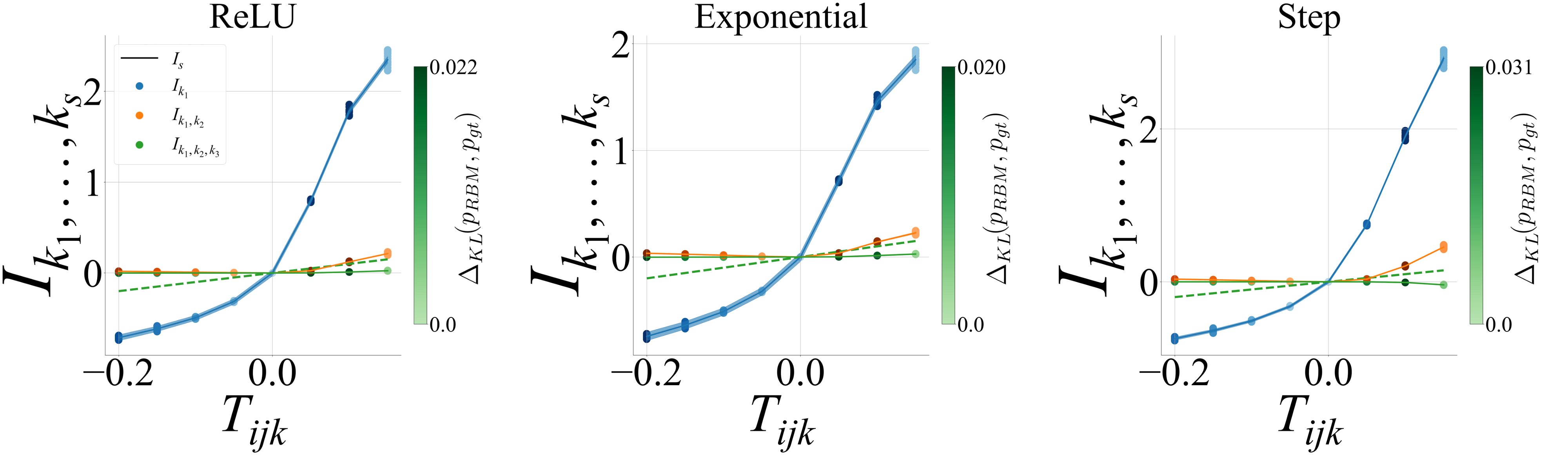}
\caption{Learning a large three-body lattice gas model. RBMs with $N = 10$ and $M =
15$, initialized with zero-mean Gaussian weights ($\sigma$ = 0.01), are trained to match the probability
distribution of ground truth lattice gas models with three-body interactions only (Eq.\eqref{three_only}) for different values of $T_{ijk}$ and for different activation functions. Each panel shows the comparison between the three-body ground truth interaction $T_{ijk}$ and the interaction terms $I_{k_1,...,k_s}$ of the trained model. The model is trained for 300 epoch with a learning rate of 0.008 (linear 0.3).}
\label{fig:NR_6_L}
\end{figure}

\begin{figure}[htbp]
\centering
\includegraphics[width=14 cm]{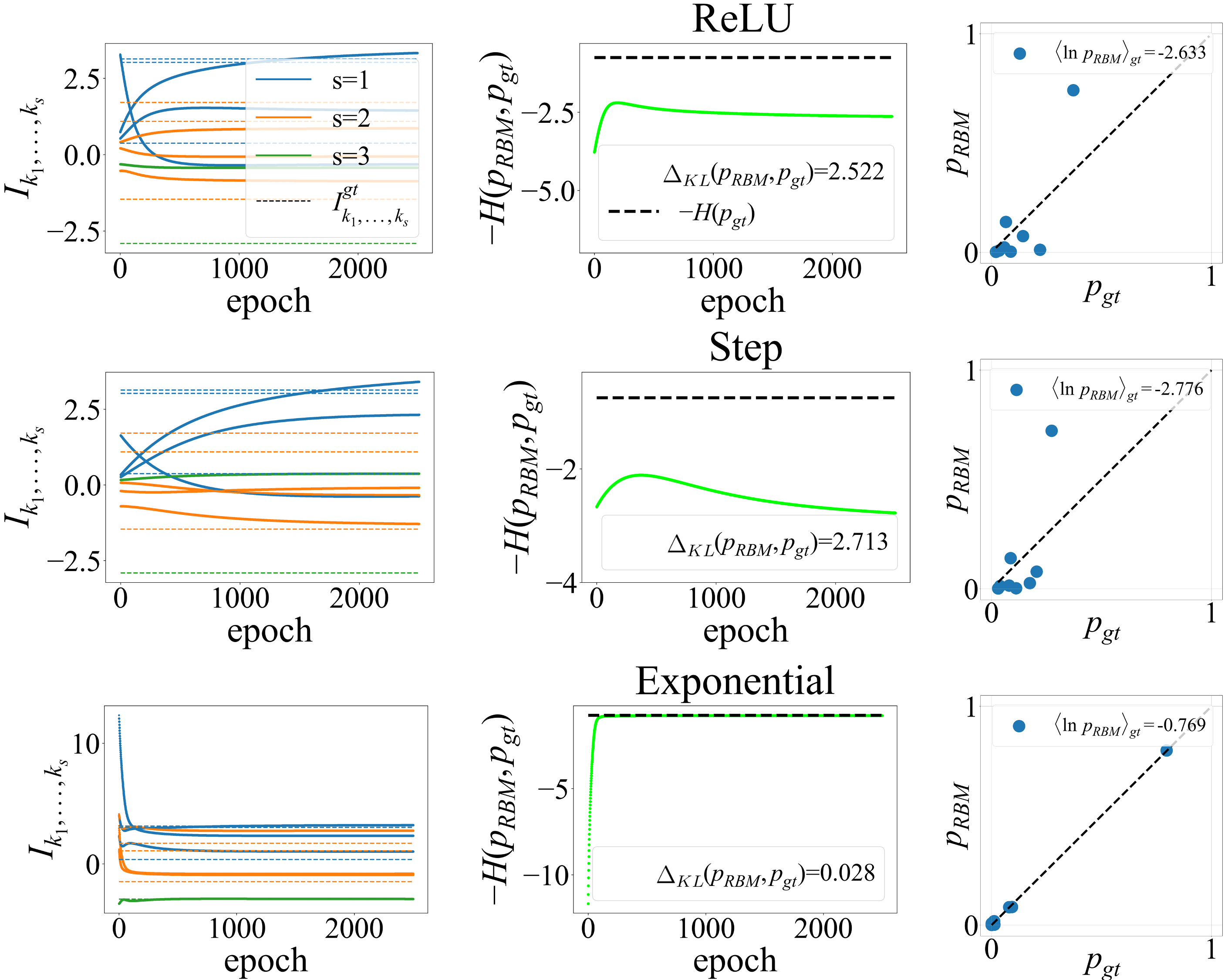}
\caption{Learning a non decaying lattice gas model with the Exponential activation - details. The model is trained for 2500 epochs with a learning rate of $5 \times 10^{-4}$. The first panel in each row shows the trajectory of the interactions mapped from the RBM, compared with the ground-truth interactions (dashed lines). The second panel in each row shows the cross-entropy trajectory, where the target is the ground truth entropy (dashed line).  $\Delta_{KL}(p_{RBM},p_{gt})$ is reported in the legend for the RBM at the end of training. The third panel in each row shows the probabilities of the states in the RBM compared to the ground truth.}
\label{fig:NR_7_supp}
\end{figure}

\newpage
\bibliographystyle{unsrt}

\end{document}